\documentclass{qjmam}

\usepackage{amssymb,amsmath,amsfonts,amsthm}
\usepackage{mathrsfs}
\usepackage{latexsym}
\usepackage{graphicx}
\usepackage{color}
\usepackage[caption=false]{subfig}
\usepackage{eepic}
\usepackage{float}

\startpage{1}
\yr{2004}
\vol{xx}
\issue{y}

\DeclareMathAlphabet\mathbit
    \encodingdefault\rmdefault\bfdefault\itdefault
\DeclareOldFontCommand{\bi}{\normalfont\bfseries\itshape}{\mathbit}

\def\reva#1{#1}
\def\rev2#1{#1}

\def\eeq{\relax}
\def\beq#1#2\eeq{\begin{equation}\label{#1}#2\end{equation}}
\def\bal#1#2\eal{\begin{align}\label{#1}#2\end{align}}
\def\bse#1#2\ese{\begin{subequations}\label{#1}#2\end{subequations}}
\renewcommand{\[}{\begin{equation}}
\renewcommand{\]}{\end{equation}}
\newcommand{\ca}{\begin{cases}}
\newcommand{\ac}{\end{cases}}
\newcommand{\ma}{\begin{pmatrix}}
\newcommand{\am}{\end{pmatrix}}

\newcommand{\bk}{\boldsymbol\kappa}

\def\fakebold#1{\relax\ifvmode\leavevmode\fi%
\ifmmode%
\setbox0=\hbox{$#1$}%
\else%
\setbox0=\hbox{#1}%
\fi%
\kern-.02em\copy0 \kern-\wd0%
\kern .04em\copy0 \kern-\wd0%
\kern-.0125em\raise.02em\box0%
}%

\def\lb{\label}
\newcommand{\er}[1]{\textrm{(\ref{#1})}}

      \def\k{\kappa}      \def\l{\lambda}     \def\m{\mu}
\def\r{\rho}        
\def\cN{{\mathcal N}}    \def\mP{{\mathscr P}}      
                  
\def\ts{\times}
\def\vp{\wedge}
\def\pa{\partial}
\def\ev{\equiv}
\def\diag{\mathop{\mathrm{diag}}\nolimits}
\def\Ker{\mathop{\mathrm{Ker}}\nolimits}

\DeclareMathOperator{\spn}{span}

\def\diag{\mathrm{diag}}

\begin{document}

\title[{wave~propagation~in~2D~and~3D~lattices}] {Wave propagation and homogenization in 2D and 3D lattices: a semi-analytical approach}

\author[a.~a.~kutsenko \etalname]{A. A. Kutsenko}

\address{Univ. Bordeaux, I2M-APY, UMR 5295, 33405 Talence, France;\\ CNRS, I2M-APY, UMR 5295, 33405 Talence, France; 
Jacobs University (International University Bremen), 28759 Bremen, Germany; Saint-Petersburg State University,
Universitetskaya nab. 7/9, St. Petersburg, 199034, Russia}

\extraauthor{A. J. Nagy \and X. Su}

\extraaddress{Mechanical and Aerospace Engineering, Rutgers University, Piscataway, NJ 08854, USA}

\extraauthor{A. L. Shuvalov}

\extraaddress{Univ. Bordeaux, I2M-APY, UMR 5295, 33405 Talence, France;\\ CNRS, I2M-APY, UMR 5295, 33405 Talence, France}

\extraauthor{A. N. Norris\footnote{$<$norris@rutgers.edu$>$}}

\extraaddress{Mechanical and Aerospace Engineering, Rutgers University, Piscataway, NJ 08854, USA}

\received{\recd x y 2016. \revd x y 2016}

\maketitle

\eqnobysec

\begin{abstract}

Wave motion in two- and three-dimensional periodic lattices of beam members supporting longitudinal and flexural waves is considered.   An  analytic method  for solving the Bloch wave spectrum is developed, characterized by a generalized eigenvalue equation  obtained by enforcing the Floquet condition.  The \rev2{dynamic stiffness} matrix is shown to be explicitly Hermitian and to admit positive eigenvalues.    Lattices with hexagonal, rectangular, tetrahedral and cubic unit cells are analyzed.   The semi-analytical method can be asymptotically expanded  for  low frequency   yielding explicit forms for the  Christoffel matrix describing wave motion in the quasistatic limit.

\end{abstract}


\section{Introduction} \label{sec1}

\rev2{Two- and three-dimensional lattices of  connected beams can provide pentamode-like behavior in the static limit
\cite{Milton95,Norris14}.  One reason for  interest in such structures  is that they exhibit one-wave behaviour characteristic of scalar or acoustic wave systems, while also displaying material anisotropy, so that anisotropic acoustic effects are possible. Such  scalar dynamic effective properties are perhaps surprising since  periodic lattice structures support 
 multiple wave types yielding complex dispersion properties} described by Bloch-Floquet spectra, particularly band gaps 
\cite{Martinsson03d,Phani06} and anisotropic propagation \cite{Gonella08}.  At the same time, the relatively simple geometry  of two- and three-dimensional lattices allows for the possibility of mechanical  simplifications that maintain the underlying structural dynamics of the continuous beam elements while leading to accurate predictions for the dispersion properties.  This paper focuses on the latter aspect, as we develop a semi-analytical formalism that reduces the Bloch wave problem to an analytically  simple form while retaining the crucial mechanics of the structure. 

\reva{Two distinct approaches \cite{WITTRICK1971} to analysing waves  in periodic structures  can be distinguished based on the number of degrees of freedom: {\it finite models} and {\it infinite models}.  The former naturally includes the 
finite elements method (FEM) for which strategies have  been  developed that are specifically designed to treat lattice structures. Thus, \cite{Phani06} developed a FEM procedure for calculating dispersion curves of hexagonal, square and triangular lattice structures.  FEM has been used by   \cite{Gonella08} to consider waves in regular and re-entrant hexagonal lattices and by   \cite{Spadoni09} to examine hexagonal chiral lattices as phononic crystals.  
Infinite methods, as in this paper, retain some of the characteristics of the continuous nature of the structure at the smallest scale.  A  simple beam  considered as a separate entity  displays an infinite number of modes; it is therefore no surprise that a model based on such elements has an infinite number of Bloch-Floquet branches.   
}

\reva{
The fundamental step in deriving the dispersion relation for  waves in a periodic structure  is the application of the Floquet condition on the unit cell of the   lattice.  Whatever approach is used, whether FEM or semi-analytical, this step reduces the problem to a generalized  eigenvalue problem for the system (or stiffness) matrix. 
The main distinction between finite and infinite models is that the former reduce to linear systems with eigenvalue equal to the square of the frequency, whereas infinite models necessarily involve finding roots of transcendental equations.  There are, however, computational approaches adapted to this problem, such as that of Wittrick and Williams \cite{WITTRICK1971} based on Householder's algorithm.  
}

Regarding other {\reva{infinite } methods for solving dynamic waves problems in lattices, we note that 
an interesting alternative wave-based approach for determining the Bloch waves in 2D periodic structures was proposed by \cite{Leamy12}.   The semi-analytical method considers the explicit waves propagating back and forth on each member, coupled by reflection and transmission matrices at joints.    The present method is similar to that of \cite{Leamy12} in that both approaches yield exact dispersion relations within the context of the beam theories employed (Timoshenko beam theory was used in \cite{Leamy12}).  However, the present approach is arguably simpler in that it does not require propagation and reflection/transmission matrices for the multiple wave types.  Instead, the crucial ingredient in the present method is the dynamic stiffness matrix that relates forces at the two ends of a beam member to the displacements at either end.


An important limit of any dynamic model is the low frequency, quasistatic or homogenization limit.  Although static homogenization theory for quite general lattice structures has been developed by several authors, e.g. \cite{Martinsson03a,Gonella08b},  these approaches do not derive the homogenized properties from  the limit of a dynamic model.  An exception is the paper by 
\cite{Colquitt11} who showed for a triangular lattice   that only by including the flexural wave effects is the effective mass properly modeled in the low frequency limit.   Simpler beam models which ignore flexural waves, or bending, show quasistatic wave speeds with effective mass that is less than the total mass of the unit cell \cite{Colquitt11}.  This suggests that models ignoring flexural effects do not properly account for the distributed mass on the wave-bearing segments of the structure, and cannot yield the correct quasistatic results.     

The analytical approach used here represents the lattice members as uni-dimensional beams supporting longitudinal and flexural waves. A strategy for implementing this was outlined by \cite{Martinsson03d} who introduced the necessary stiffness matrix relating forces and displacements at the ends of a beam.  By combining these matrices it is possible to represent any periodic lattice, in principle. The method of \cite{Martinsson03d} was used in \cite{Colquitt11} to consider lattices with triangular unit cell structure, and for square cell lattices in \cite{Colquitt13}. 
In this paper we develop further the approach proposed by \cite{Martinsson03d} and \cite{Colquitt11}.  We present, for the first time,  analysis of a general  hexagonal unit cell lattice, a structure  of great interest in relation to graphene and other phenomena. Also, the method is extended into 3D to analyze the tetrahedral unit cell lattice. 
\reva{The formulation is semi-analytical to the extent that all matrix elements are explicit, the dispersion relation for square and cubic lattices are derived analytically. Although one could obtain analytic dispersion relations for hexagonal and tetrahedral lattices using symbolic computation \cite{Williams1995}, direct numerical methods are employed at the final stage to perform  computation}. The semi-analytical nature of the solution allows us to extract the low frequency asymptotics, and to find closed-form expressions for the quasistatic Christoffel matrix, as demonstrated  for hexagonal and rectangular unit cell lattices in 2D. 
\rev2{In this sense the present study is  step in the continuation from low frequency (quasi-static) response governed by effective elastic stiffness and density to dynamic effective medium models.  }

\rev2{   The present analysis does not include torsion in the individual members.  The beams are assumed to have large length to thickness ratio, and hence a static applied  macroscopic torsion is borne at the level of the unit cell by flexure of the members.  Bending is the dominant effect for producing torsion in the lattice structures considered here. This can be seen {\it a posteriori} from the comparisons below with full elastodynamic simulations which do not display Bloch-Floquet branches with significant torsional effects at level of the lattice member.  In other words, torsion in individual members is ignored because we are only including the dynamic counterparts of the micro-effects that lead to the static effective medium. Note that the present model allows for rigid body rotation at the unit cell level, which is consistent with static homogenization \cite{Norris14}. 
}

%

The format of the paper is as follows.  The solution method  for hexagonal and tetrahedral lattices is summarized in \S \ref{sec2}, where the Bloch wave condition is explicitly used to derive the dispersion relation for Floquet modes. The detailed derivation of the system matrix for the hexagonal lattice is presented in \S \ref{sec3}.  The low frequency asymptotics are examined in \S \ref{sec4} where  the explicit form of the quasi-static Christoffel matrix is derived.   The dynamic and quasi-static solutions are obtained for the rectangular lattice in \S \ref{sec5}, and for cubic lattice in \S \ref{sec6}.  In addition, numerical examples in \S \ref{sec5} and \S \ref{sec6} compare  results from the present theory with fully elastodynamic FEM computations for hexagonal, rectangular, tetrahedral and cubic lattices. 

\section{\rev2{Dispersion relation}} \label{sec2}

\subsection{\rev2{Structures and structural parameters}}

We focus our attention on two example   structures in 2D and 3D,  hexagonal and tetrahedral lattices, respectively.  Each may be defined by two points ${\bf a}_1$, ${\bf a}_2$ inside the unit cell $\mP$ spanned by  vectors ${\bf e}_1$, ${\bf e}_2$ (and ${\bf e}_3$ in 3D), see Fig.\ \ref{fig1}. The unit cell $\mP$ 
is then periodically translated to cover the whole plane (space in 3D) and thereby make the infinitely extended lattice. 
We assume that all material parameters are periodic such that the properties in any translated cell $\mP+n{\bf e}_1+m{\bf e}_2\ \ (+l{\bf e}_3\ \ {\text {in 3D}})$  coincide with those in $\mP$.

Every point ${\bf
a}_i$ in the lattice is connected to three (four in 3D) neighboring points ${\bf a}_j$ by rods
$[{\bf a}_i,{\bf a}_j]$ with length $l_{ij}=|{\bf a}_i-{\bf a}_j|$
and direction ${\bf e}_{ij}=l_{ij}^{-1}({\bf a}_i-{\bf a}_j)$.
There are  masses $m_1$, $m_2$ with moments of inertia $I_1$, $I_2$ at the points ${\bf a}_1$, ${\bf a}_2$.
The rod $[{\bf a}_i,{\bf a}_j]$ has axial stiffness $\m_{ij}$,   beam flexural coefficient $\l_{ij}$ and lineal density $\r_{ij}$ (these are related to the rod Young's modulus $E_{ij}$,  cross-sectional area $A_{ij}$, radius of gyration $\kappa_{ij}$ and volumetric density $\rho_{ij}^V$ by
$\m_{ij} =E_{ij}A_{ij}$, $\l_{ij} =E_{ij}A_{ij}\kappa_{ij}^2$, $\r_{ij} = \rho_{ij}^VA_{ij}$).

\begin{figure}[ht]
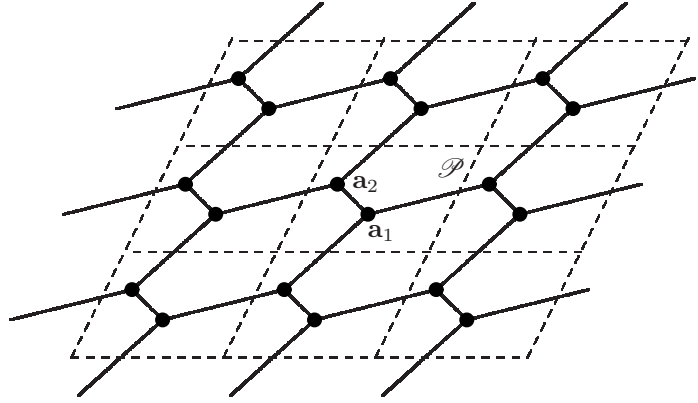

\centering
\setlength{\unitlength}{1mm}

                         \caption{The hexagonal (honeycomb) lattice}
{\label{fig1}}
                           \end{figure}
\begin{figure}[ht]
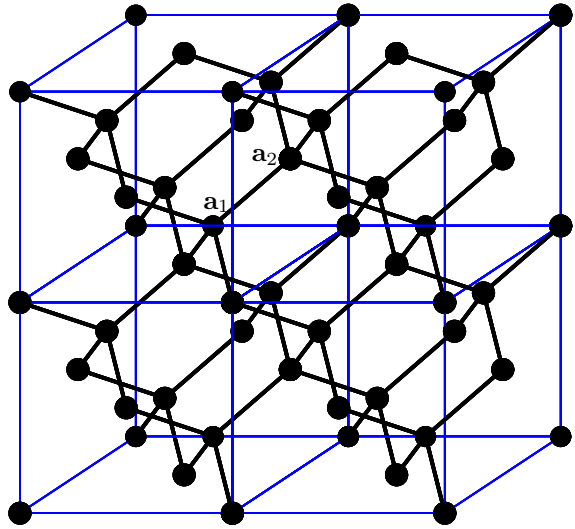

\centering
\setlength{\unitlength}{0.254mm}
%
 \caption{The tetrahedral lattice}
         \label{diamond}
\end{figure}

\subsection{Analytic dispersion relation}

Considering the rod ${\bf e}_{ij}$, let ${\bf u}_i$, ${\bf u}_j$ denote the displacement at the end  points ${\bf a}_i$, ${\bf a}_j$, respectively.  Let  ${\bf f}_{ij}$ denote the force at point ${\bf a}_i$ from the rod ${\bf e}_{ij}$.  The precise form of the displacement and force 3-vectors (6-vectors for 3D case) will be defined in Section \ref{sec3}, for the moment we do not need to know their specific nature, except to note that they include both longitudinal and flexural effects.   The equilibrium  equation at point ${\bf a}_i$  is then
\beq{8} \sum\limits_{j\in \mathcal{N}_i} {\bf  f}_{ij} = -\omega^2
{\bf M}_i {\bf u}_i, \quad
{\bf M}_i = \diag ( m_i,  m_i, I_i) \ \ \text {in 2D or }  \diag ( m_i,  m_i,   m_i, I_i, I_i, I_i) \ \ \text {in 3D}, 
\eeq
where  $\cN_i$ is the set of points  connected with ${\bf a}_i$. \reva{It is notable that this approach allows concentrated masses at the junctions which are included in the matrix ${\bf M}_i$.}
The force ${\bf f}_{ij}$  may be expressed in terms of the end point displacements
\beq{17} {\bf f}_{ij}  =
{\bf P}_{ij}^{(2)} {\bf u}_j - {\bf P}_{ij}^{(1)} {\bf u}_i ,
\eeq
where the frequency dependent stiffness matrices ${\bf P}_{ij}^{(1)} (\omega)$, ${\bf P}_{ij}^{(2)} (\omega)$ are derived in Section \ref{sec3}.

Applying the Floquet periodic conditions
\[\lb{221}
\begin{aligned}
 {\bf u}_j&=\exp(i{\bf k}\cdot{\bf g}_{j}){\bf u}_1,\ \ {\bf g}_j={\bf
 a}_j-{\bf a}_1,\ \ j\in\cN_2 ,
\\ 
 {\bf u}_j&=\exp(i{\bf k}\cdot{\bf g}_{j}){\bf u}_2,\ \ {\bf g}_j={\bf
 a}_j-{\bf a}_2,\ \ j\in\cN_1
\end{aligned}
\]
and using eqs.\ \er{8} and \er{17} leads to
\beq{3-3}
\begin{aligned}
\sum\limits_{j\in
\mathcal{N}_1} \big(
 {\bf P}_{1j}^{(2)} \exp(i{\bf k}\cdot{\bf g}_{j}) \, {\bf u}_2
- {\bf P}_{1j}^{(1)} {\bf u}_1 \big)
&= -\omega^2 {\bf M}_1 {\bf
u}_1 ,
\\ 
\sum\limits_{j\in \mathcal{N}_2} \big(
 {\bf P}_{2j}^{(2)} \exp(i{\bf k}\cdot{\bf g}_{j}) \, {\bf u}_1
- {\bf P}_{2j}^{(1)} {\bf u}_2 \big)
&= -\omega^2 {\bf M}_2 {\bf
u}_2.
\end{aligned}
\eeq
For each $j\in \mathcal{N}_2$ there is a unique $\bar j\in \mathcal{N}_1$ such that
\beq{012}
{\bf P}_{2j}^{(2)} e^{i {\bf k}\cdot{\bf g}_j}
=
\Big( {\bf P}_{1\bar j}^{(2)} e^{i {\bf k}\cdot{\bf g}_{\bar j}}
\Big)^{+},
\ \
{\bf P}_{2j}^{(1)} = {\bf P}_{1\bar j}^{(3)},
\ \
\eeq
where $+$ denotes the Hermitian conjugation and the matrices $ {\bf P}_{1j}^{(3)}$ are defined in Section \ref{sec3}.  Hence it is possible to express the second equation of \eqref{3-3} in terms of a sum over neighboring links of ${\bf a}_1$.
Introducing matrices
\[\lb{223}
 {\bf H}_1=\sum\limits_{j\in\mathcal{N}_1}
 {\bf P}_{1j}^{(1)},\ \  {\bf H}_2=-\sum\limits_{j\in\mathcal{N}_1}
 {\bf P}_{1j}^{(2)}\exp(i{\bf k}\cdot{\bf g}_{j})
,\ \ {\bf H}_3=\sum\limits_{j\in\mathcal{N}_1}{\bf P}_{1j}^{(3)}
\]
equations \er{3-3} can then be rewritten in the form
\[\lb{224}
 {\bf H}{\bf u}=\omega^2{\bf M}{\bf u}
\]
with
\[\lb{225}
 {\bf u}=\ma {\bf u}_1 \\ {\bf u}_2 \am,\ \ {\bf M}=\diag({\bf M}_1,{\bf
 M}_2),\ \ {\bf H}\ev{\bf H}(\omega,{\bf k})=\ma {\bf H}_1 & {\bf H}_2 \\ {\bf H}_2^+ & {\bf
 H}_3 \am  \ \ \big(  = {\bf H}^+\big).
\]
Then Floquet curves (dispersion curves) $\omega_n({\bf k})$ can be found
from the equation
\[\lb{226}
 \det({\bf H}(\omega,{\bf k})-\omega^2{\bf M})=0.
\]
Note that according to Section \ref{sec3} (see eqs.\ \eqref{223}, \eqref{10}, \eqref{003} and \eqref{3}), the matrices ${\bf H}_1$ and ${\bf H}_3$ are real symmetric, so that the  matrix   $ {\bf H}$
is Hermitian, in turn guaranteeing that the dispersion relation
\er{226} is real valued for real $\omega$, ${\bf k}$.  We will return to
this equation in Section \ref{sec3} after we have described the
displacements and forces, and derived the stiffness matrices.

\section{Dynamic stiffness matrices} \label{sec3}

\subsection{Longitudinal wave equation} Consider the rod ${\bf
e}_{ij}$ with uniform Young's modulus $\mu_{ij}$ and density $\r_{ij}$.
Let $u_{ij}(x)$ denote the component of the displacement in the ${\bf e}_{ij}-$direction at any point $x$ (a one-dimensional linear coordinate parameter) of $[{\bf a}_i,{\bf a}_j]$. The displacement $u_{ij}$ satisfies the  wave equation for longitudinal wave motion and its associated   boundary
conditions (BCs)
\[\lb{201}
 \m_{ij}\frac{\pa^2 }{\pa x^2}u_{ij}=-\omega^2\r_{ij}u_{ij},\ \ u_{ij}(0)={\bf
 e}_{ij}\cdot{\bf u}_{i},\ \ u_{ij}(l_{ij})={\bf e}_{ij}\cdot{\bf
 u}_j.
\]
Solving \er{201}, 
\[\lb{203}
 u_{ij}(x)=\frac{
{\bf e}_{ij}\cdot{\bf  u}_{i}\sin(s_{ij}\omega (l_{ij}-x))
+
{\bf e}_{ij}\cdot{\bf u}_{j} \sin(s_{ij}\omega  x)}{\sin(s_{ij}\omega l_{ij})},
\ \ s_{ij}=\sqrt{\frac{\r_{ij}}{\m_{ij}}},
\]
implying that the longitudinal force ${\bf f}_{ij}$ acting on the point ${\bf a}_i$ is
\[\lb{204}
 {\bf f}_{ij\ wave} \equiv 
\m_{ij}\frac{\pa u_{ij} }{\pa x}(0)\, {\bf e}_{ij}
=\frac{  \m_{ij}s_{ij}\omega}{\sin(s_{ij}\omega l_{ij})}\, 
{\bf e}_{ij}{\bf e}_{ij}^{T} 
\big( {\bf u}_{j}-{\bf
 u}_{i}\cos(s_{ij}\omega l_{ij})\big)
 .
\]

\subsection{Flexural wave equation} 

\rev2{The kinematic BCs for flexural wave motion involve both the flexural displacement and the non-torsional rotation at the ends of the rod.   In 2D, define the unit vector perpendicular to the plane of the lattice, ${\bf e}_b= 
{\bf e}_1\vp {\bf e}_2 /|{\bf e}_1\vp {\bf e}_2|$.  
The flexural displacement $v_{ij}(x)$  at any point $x$ on the rod ${\bf e}_{ij}$ is then defined as  the component of the displacement in the ${\bf e}_{ij}^\perp-$ direction, 
where 
}
\beq{6} 
{\bf e}_{ij}^\perp =  { {\bf e}_{b}  \vp {\bf e}_{ij} }. 
\eeq
\rev2{The generalized 2D displacement vectors are therefore ``three-dimensional" with two  components for the longitudinal motion and one  for flexural. 
The flexural wave equation and its BCs are, with  $v ' = \partial v /\partial x$, }
\bal{4}
 & - \l_{ij} \frac{\partial^4
v_{ij}}{\partial x^4}   =- \omega^2 \r_{ij} v_{ij} , \quad
\\
v_{ij}(0)= {\bf e}_{ij}^\perp \cdot {\bf u}_i, \ \  &
v_{ij}(l_{ij})= {\bf e}_{ij}^\perp \cdot{\bf u}_j, \ \ v_{ij}'(0)=
{\bf e}_b \cdot{\bf u}_i, \ \ v_{ij}'(l_{ij})= {\bf e}_b\cdot {\bf
u}_j . \nonumber \eal
The generalized force (shear force and bending moment) at point ${\bf a}_i$ due to bending is \beq{7} {\bf f}_{ij\ bending}  = -\lambda_{ij}
\frac{\partial^3 v_{ij}}{\partial x^3} {\bf e}_{ij}^\perp +
\lambda_{ij} \frac{\partial^2 v_{ij}}{\partial x^2} {\bf e}_b. \eeq

\rev2{In 3D we extend the definition of the  end point flexural displacement  by defining   two  non-torsional  rotation components in the directions ${\bf e}_b$ and ${\bf e}_b^\prime$, and the related shear force   components along 
${\bf e}_{ij}^\perp $ and ${\bf e}_{ij}^{\perp \prime}$, where }
\beq{new perp}
{\bf e}_{ij}^\perp = ({\bf r}, {\bf 0}_3),\ \ {\bf e}_b = ({\bf 0}_3, {\bf e}_{ij} \vp {\bf r}),\ \ {\bf e}_{ij}^{\perp \prime} = ({\bf e}_{ij} \vp {\bf r}, {\bf 0}_3),\ \ {\bf e}_b^\prime = -({\bf 0}_3, {\bf r}) .
\eeq
Here ${\bf 0}_3$ is 3D zero-vector, ${\bf r} = {\bf e}_{\alpha} \vp {\bf e}_{ij}/|{\bf e}_{\alpha} \vp {\bf e}_{ij}|$ and ${\bf e}_\alpha$ is some (any) vector ${\bf e}_1$ or ${\bf e}_2$ or ${\bf e}_3$ whichever is not parallel to ${\bf e}_{ij}$. 
\rev2{Let $w_{ij}(x)$ denote the displacement in direction ${\bf e}_{ij}^{\perp \prime}$, $w_{ij} ' = \partial w_{ij} /\partial x$, 
then the flexural wave equation and its BCs for the extra dimension in 3D case can be written as eq.\ \eqref{4} combined with }
\bal{42}
 & - \l_{ij} \frac{\partial^4
w_{ij}}{\partial x^4}   =- \omega^2 \r_{ij} w_{ij} , \quad
\\
w_{ij}(0)= {\bf e}_{ij}^{\perp\prime} \cdot {\bf u}_i, \ \  &
w_{ij}(l_{ij})= {\bf e}_{ij}^{\perp\prime}  \cdot{\bf u}_j, \ \ w_{ij}'(0)=
{\bf e}_b^\prime \cdot{\bf u}_i, \ \ w_{ij}'(l_{ij})= {\bf e}_b^\prime \cdot {\bf
u}_j . \nonumber \eal
The additional generalized force term at point ${\bf a}_i$ is 
\beq{7_2} {\bf f}^{\prime}_{ij\ bending}  = -\lambda_{ij}
\frac{\partial^3 w_{ij}}{\partial x^3} {\bf e}_{ij}^{\perp \prime} +
\lambda_{ij} \frac{\partial^2 w_{ij}}{\partial x^2} {\bf e}_b^\prime. \eeq

\rev2{In summary, the    components of $ {\bf u}_i$
in the ${\bf e}_{ij}^\perp$, ${\bf e}_{ij}^{\perp\prime}$ and ${\bf e}_b$, ${\bf e}_b^\prime$ directions are the transverse deflection and  beam rotation angle, respectively.
The force components in direction ${\bf e}_{ij}^\perp$ and ${\bf e}_{ij}^{\perp\prime}$ are the  resultant shear force while   the
${\bf e}_b$ and ${\bf e}_b^\prime$ "force" components represent the bending moment. Note that in 3D, there are three displacement components and three rotation components.
In this way the coupled longitudinal and flexural dynamics of the 2D lattice  are described in terms of  "three-dimensional"vectors for  displacement and forces in 2D, and "six-dimensional" vectors for 3D lattices.}

The generalized forces at the two ends of the rod are related  to the
displacements there by the stiffness matrix ${\bf K}$,  defined such that
\bal{003}
 \begin{pmatrix} {\bf e}_{ij}^{\perp}\cdot {\bf f}_{ij} \\ {\bf e}_b\cdot{\bf f}_{ij}\\ {\bf e}_{ij}^{\perp}\cdot {\bf f}_{ji} \\ {\bf e}_b\cdot{\bf f}_{ji} \end{pmatrix}
 = -\l_{ij}
 {\bf K}(\omega)
\begin{pmatrix} {\bf e}_{ij}^{\perp}\cdot {\bf u}_i \\ {\bf e}_b\cdot{\bf u}_i \\ {\bf e}_{ij}^{\perp}\cdot {\bf u}_j \\ {\bf e}_b\cdot{\bf u}_j \end{pmatrix}
,
\quad {\bf K} =
\begin{pmatrix}
 {\bf K}_1 &  {\bf K}_2
\\
 {\bf K}_2^T &  {\bf K}_3
\end{pmatrix}.
 \eal
The bending forces \eqref{7} and \eqref{7_2} at  lattice site  $i$ from rod $ij$ therefore becomes
\beq{7a}
\begin{aligned}
 &{\bf f}_{ij\ bending}  = -\lambda_{ij}\big( {\bf e}_{ij}^\perp  , \, {\bf e}_b\big)\Big( {\bf
 K}_1
\big( {\bf e}_{ij}^\perp  , \, {\bf e}_b\big)^{T}{\bf u}_i+ {\bf
 K}_2
\big( {\bf e}_{ij}^\perp  , \, {\bf e}_b\big)^{T}{\bf u}_j\Big), 
\\
 &{\bf f}^{\prime}_{ij\ bending}  = -\lambda_{ij}\big( {\bf e}_{ij}^{\perp \prime}  , \, {\bf e}_b^\prime \big)\Big( {\bf
 K}_1
\big( {\bf e}_{ij}^{\perp \prime}  , \, {\bf e}_b^\prime \big)^{T}{\bf u}_i+ {\bf
 K}_2
\big( {\bf e}_{ij}^{\perp \prime}  , \, {\bf e}_b^\prime \big)^{T}{\bf u}_j\Big).
\end{aligned}
\eeq
We next derive the explicit form of the stiffness matrix. 

\subsection{Solution of the flexural stiffness matrix}
With eqs.\ \er{4} and \er{42} in mind, consider the
solution to
\beq{1} \frac{\partial^4 w}{\partial x^4} - \gamma^4
w=0, \quad x\in[0,l], \eeq in the form \bal{2} &w(x) = \frac 1{2(1-
c c_h)} \big\{ \big[ ( c -c_h )(\cos \gamma x-\cosh \gamma x) + (s+
s_h)(\sin \gamma x -\sinh \gamma x )\big] w(l)
\nonumber \\
& + \frac 1{\gamma}\big[ ( s_h-s )(\cos \gamma x-\cosh \gamma x) +
(c-c_h)(\sin \gamma x -\sinh \gamma x )\big] w'(l)
\nonumber \\
& + \big[ (1-c c_h +s s_h)\cos \gamma x +(1-c c_h -s s_h)\cosh
\gamma x + (c s_h+s c_h)(\sinh \gamma x -\sin \gamma x )\big] w(0)
\nonumber \\
& + \frac 1{\gamma} \big[ (s c_h-c s_h)(\cos \gamma x -\cosh \gamma
x )+ (1-c c_h -s s_h)\sin \gamma x +(1-c c_h +s s_h)\sinh \gamma x
\big] w'(0) \big\},
\eal
where $c=\cos \gamma l$, $s=\sin \gamma l$,  $c_h=\cosh \gamma l$,
$s_h=\sinh \gamma l$. $v(x)$ and $w(x)$ have the same form of solution, so that the stiffness matrix is the same. According to its definition in \eqref{003} the stiffness matrix ${\bf K}$ satisfies 
\bal{3=5}
\begin{pmatrix}
w'''(0)
\\
-w''(0)
\\
-w'''(l)
\\
w''(l)
\end{pmatrix}
&=  {\bf K}(\omega)
\begin{pmatrix}
w(0)
\\
w'(0)
\\
w(l)
\\
w'(l)
\end{pmatrix} .
\eal
The explicit form of the stiffness matrix  then follows from \eqref{2}  as 
\bal{3}
 {\bf K}(\omega)
= \frac {\gamma^2}{1- c c_h}
\begin{pmatrix}
\gamma(cs_h+sc_h)  &  ss_h & -\gamma (s+s_h) &
c_h-c
\\
 s s_h & \gamma^{-1} (sc_h - c s_h ) &  c-c_h & \gamma^{-1}
(s_h -s)
\\
-\gamma (s+s_h)  &  c- c_h & \gamma(cs_h+sc_h) &
-  ss_h
\\
 c_h-c  &  \gamma^{-1} (s_h-s) & -  s s_h & \gamma^{-1}
(sc_h-cs_h)
\end{pmatrix}.
\eal

\subsection{Total force and stiffness matrices }
The total force at point $i$ from rod ${\bf e}_{ij}$  now follows from 
\eqref{204} and \eqref{7}, 
\beq{71} {\bf f}_{ij
} = {\bf f}_{ij\ wave}(0) + {\bf f}_{ij\ bending}(0) + {\bf f}^{\prime}_{ij\ bending}(0),
\eeq
where ${\bf f}^{\prime}_{ij\ bending}(0)$ doesn't exist in 2D case. Set
\bal{23} &\tilde{\mu}_{ij} = {\mu}_{ij}/l_{ij}, \quad 
\tilde{s}_{ij} (\omega) = \omega s_{ij} l_{ij}, \quad
\gamma_{ij} (\omega) =\big(\omega^2 \r_{ij}/\l_{ij} \big)^{1/4} ,\quad
{\bf A}_{ij} =  {\bf e}_{ij}
{\bf e}_{ij}^T,
\eal
The dynamic stiffness matrices introduced in eqs.\ \er{17} and \er{012} then follow from
\er{204}, \er{7a} and \er{71} as
\bal{10} {\bf P}_{ij}^{(1)} & =
\tilde\mu_{ij} \tilde s_{ij}\cot \tilde s_{ij}   {\bf
A}_{ij} +\lambda_{ij}
 \big( {\bf e}_{ij}^\perp  , \, {\bf e}_b\big) {\bf K}_1
\big( {\bf e}_{ij}^\perp  , \, {\bf e}_b\big)^{T} +\lambda_{ij}
 \big( {\bf e}_{ij}^{\perp \prime}  , \, {\bf e}_b^\prime \big) {\bf K}_1
\big( {\bf e}_{ij}^{\perp \prime}  , \, {\bf e}_b^\prime \big)^{T} ,
\nonumber \\
{\bf P}_{ij}^{(2)}&= \tilde\mu_{ij}\tilde s_{ij} \csc \tilde
s_{ij}  {\bf A}_{ij} - \lambda_{ij}
 \big( {\bf e}_{ij}^\perp  , \, {\bf e}_b\big) {\bf K}_2
\big( {\bf e}_{ij}^\perp  , \, {\bf e}_b\big)^{T} - \lambda_{ij}
 \big( {\bf e}_{ij}^{\perp \prime}  , \, {\bf e}_b^\prime \big) {\bf K}_2
\big( {\bf e}_{ij}^{\perp \prime}  , \, {\bf e}_b^\prime \big)^{T} ,
\\
{\bf P}_{ij}^{(3)} & = \tilde\mu_{ij} \tilde s_{ij}\cot
\tilde s_{ij}   {\bf A}_{ij} +\lambda_{ij}
 \big( {\bf e}_{ij}^\perp  , \, {\bf e}_b\big) {\bf K}_3
\big( {\bf e}_{ij}^\perp  , \, {\bf e}_b\big)^{T} +\lambda_{ij}
 \big( {\bf e}_{ij}^{\perp \prime}  , \, {\bf e}_b^\prime \big) {\bf K}_3
\big( {\bf e}_{ij}^{\perp \prime}  , \, {\bf e}_b^\prime \big)^{T}.   \nonumber \eal
where ${\bf K}$ is defined by  eq.\ \eqref{3} with  
\beq{-12} 
\gamma =   \gamma_{ij}  , \ \ 
c=\cos \gamma_{ij}l_{ij},\ \ s=\sin \gamma_{ij}l_{ij},\ \ c_h=\cosh \gamma_{ij}l_{ij},\ \ s_h=\sinh \gamma_{ij}l_{ij} .
\eeq 
The identities \eqref{012}  are a consequence of the relations
${\bf K}_2^T = {\bf J}{\bf K}_2{\bf J}$, ${\bf K}_3 = {\bf J}{\bf K}_1{\bf J}$
where  ${\bf J} = \diag(1,-1)$. 
The force  ${\bf f}_{ij}$ at point ${\bf a}_i$ given by eq.\ \eqref{17} 
then follows from \eqref{204},  \eqref{7a} and \eqref{71}.

\section{Effective wave speeds at low frequency} \label{sec4}

\subsection{Low frequency asymptotics}

The  low-frequency asymptotic behavior of ${\bf K}$ defined in
\er{3} is, using ${\bf K}_3 = {\bf J}{\bf K}_1{\bf J}$, 
\beq{45}
\begin{aligned}
{\bf K}_1  &= l^{-2}
\begin{pmatrix} {12}\,l^{-1} & 6
\\
6 & {4}\,{l}
\end{pmatrix} + \frac{\gamma^4 l^2}{35}
 \begin{pmatrix}
-13\,l^{-1} & -\frac{11}6
\\
-\frac{11}6  & -\frac{l}3
\end{pmatrix} + \text{O}( \gamma^8) ,
\\
{\bf K}_2  &=  l^{-2}\begin{pmatrix} -{12}\,l^{-1} & 6
\\
-6 & 2\,l
\end{pmatrix} + \frac{\gamma^4 l^2}{70}
 \begin{pmatrix}
-9\,l^{-1}  & \frac{13}6
\\
-\frac{13}6  & \frac{l}2
\end{pmatrix} + \text{O}( \gamma^8)  ,
\end{aligned} 
\eeq
 implying  that    ${\bf K}(0)$ is  positive semi-definite
having  eigenvalues $30$ and $2$ with non-normalized eigenvectors
$(2,1,-2,1)^T$ and $(0,1,0,-1)^T$, respectively. The null vectors of
${\bf K}(0)$, $(1,0,1,0)^T$ and $(-l,2,l,2)^T$, correspond to rigid
body displacement and rotation, respectively.

The low frequency expansions of the dynamic stiffness matrices of eq. \eqref{10} are
\beq{11}
\begin{aligned}
{\bf P}_{ij}^{\big(\stackrel{1}{3} \big)} (\omega )=& \tilde\mu_{ij}{\bf
A}_{ij}  + 2
 \lambda_{ij} l_{ij}^{-3} \Big(
 6({\bf A}_{ij}^{\perp  }+{\bf A}_{ij}^{\perp  \prime})
 \pm 3({\bf A}_{ij}^{b\perp  }+{\bf A}_{ij}^{b\perp  \prime} 
 +{\bf A}_{ij}^{\perp b }+{\bf A}_{ij}^{\perp b \prime})l_{ij}  
+  2 ({\bf A}_b+{\bf A}_b^\prime)l_{ij}^2
\Big)
\\
& -  \frac 13 \omega^2 \r_{ij} l_{ij} \Big(  {\bf A}_{ij} +
 \frac{1}{70}   \big(
 78({\bf A}_{ij}^{\perp  }+{\bf A}_{ij}^{\perp  \prime})
\pm 11({\bf A}_{ij}^{\perp b }+{\bf A}_{ij}^{\perp b \prime}   
 +{\bf A}_{ij}^{b\perp  }+{\bf A}_{ij}^{b\perp  \prime})l_{ij}\\
& +  2
({\bf A}_b+{\bf A}_b^\prime)l_{ij}^2\big) \Big) +\text{O}(\omega^4),
\\
{\bf P}_{ij}^{(2)} (\omega )=& \tilde\mu_{ij}{\bf A}_{ij}  + 2\lambda_{ij}l_{ij}^{-3}\Big(
 6({\bf A}_{ij}^{\perp  }+{\bf A}_{ij}^{\perp  \prime})
+ 3({\bf A}_{ij}^{b\perp  }+{\bf A}_{ij}^{b\perp  \prime}
     - {\bf A}_{ij}^{\perp b }-{\bf A}_{ij}^{\perp b \prime})l_{ij} 
-  ({\bf A}_b+{\bf A}_b^\prime)l_{ij}^2
\Big)
\\ & + \frac 16 \omega^2 \r_{ij} l_{ij} \Big( {\bf A}_{ij} +
 \frac{1}{70}   \big(
 54({\bf A}_{ij}^{\perp  }+{\bf A}_{ij}^{\perp  \prime})
+ 13({\bf A}_{ij}^{b\perp  }+{\bf A}_{ij}^{b\perp  \prime}
      -{\bf A}_{ij}^{\perp b }-{\bf A}_{ij}^{\perp b \prime} )l_{ij}
\\
& -  3
({\bf A}_b+{\bf A}_b^\prime)l_{ij}^2 \big) \Big)  +\text{O}(\omega^4) ,
\end{aligned}
\eeq
where 
\bal{62}
&{\bf A}_{ij}^\perp = {\bf e}_{ij}^\perp {{\bf
e}_{ij}^\perp  }^T, \ \ \quad {\bf A}_{ij}^{\perp b} =  {\bf
e}_{ij}^\perp  {\bf e}_b^T, \ \ \quad {\bf A}_{ij}^{b\perp  } =
{\bf e}_b  {{\bf e}_{ij}^\perp   }^T, \ \ \ \ {\bf A}_b =  {\bf e}_b
{\bf e}_b^T , \\
&{\bf A}_{ij}^{\perp \prime} = {\bf e}_{ij}^{\perp \prime} {{\bf
e}_{ij}^{\perp \prime}  }^T, \quad {\bf A}_{ij}^{\perp b \prime} =  {\bf
e}_{ij}^{\perp \prime}  {{\bf e}_b^{\prime}}^T, \quad {\bf A}_{ij}^{b\perp  \prime} =
{\bf e}_b^\prime  {{\bf e}_{ij}^{\perp \prime}   }^T, \quad {\bf A}_b^\prime =  {\bf e}_b^\prime
{{\bf e}_b^\prime}^T . \label{62_2}
\eal
Note that the terms with the primes are not present for the 2D lattice, and hence  eq.\ \eqref{62_2} applies only for the 3D case.
The zero frequency limit of the system matrix  ${\bf H}$ defined in eq.\ \eqref{225}  has the following form
\[\lb{400}
{\bf H}^{(0)}\equiv {\bf H}(0,{\bf 0}) =\ma {\bf H}_+^{(0)} & -{\bf H}_+^{(0)} \\ -{\bf H}_-^{(0)} & {\bf H}_-^{(0)} \am
+\ma {\bf 0} & {\bf R}_+ \\ {\bf R}_- & {\bf 0} \am
\]
with
\beq{401}
\begin{aligned}
 {\bf H}_\pm ^{(0)}&=\sum_{j\in\cN_1}\Big(
\tilde\mu_{1j}{\bf A}_{1j}  + 2  \lambda_{1j}l_{1j}^{-3} \big(
 6({\bf A}_{ij}^{\perp  }+{\bf A}_{ij}^{\perp  \prime})
 \pm 3({\bf A}_{ij}^{b\perp  }+{\bf A}_{ij}^{b\perp  \prime} 
  +{\bf A}_{ij}^{\perp b }+{\bf A}_{ij}^{\perp b \prime})l_{ij}  
	+  2 ({\bf A}_b+{\bf A}_b^\prime)l_{ij}^2
\big)
\Big),
\\ 
 {\bf R}_\pm &=\sum_{j\in\cN_1}6\l_{1j}l_{1j}^{-3}\Big( ({\bf A}_b+{\bf A}_b^\prime )l_{ij}^2 \pm 2({\bf A}_{1j}^{\perp b}+{\bf A}_{1j}^{\perp b \prime})l_{ij} \Big).
\end{aligned}
\eeq

 The effective quasi-static
speeds are defined as
\[\lb{408}
 c(\pmb{\k})=\lim_{k\to0}\frac{\omega({\bf k})}{k},\ \ {\bf  k}=k\pmb{\k},\ \ |\pmb{\k}|=1.
\]
We consider the  following  perturbation {\it ansatz}  for small
${\bf k}$,  
\[\lb{410}
 \omega^2({\bf k})=  k\omega_1+
k^2\omega_2+O(k^3)
\]
with associated displacement 
\[\lb{411}
 {\bf u}({\bf k})={\bf u}^0+k{\bf u}^{1}+k^2{\bf u}^2+{\bf O}(k^3) .
\]
The  asymptotic behavior of ${\bf H}$ for small $\omega$ and $k$ is
\[\lb{409}
 {\bf H}(\omega,{\bf k})={\bf H}^{(0)}+k{\bf H}^{(1)}(\pmb{\k})+
 k^2{\bf H}^{(2)}(\pmb{\k})+\omega^2{\bf H}^{(3)}+{\bf O}(\omega^4)+{\bf O}(k\omega^2)+{\bf O}(k^3).
\]
Substituting \er{411}-\er{409} into \er{224} and identifying  terms with
 the same power of $k$, 
yields  at O$(k)$
\[\lb{412}\
{\bf H}^{(0)} {\bf u}^1 +{\bf H}^{(1)}{\bf u}^0
= \omega_1 ( {\bf M}-{\bf  H}^{(3)}){\bf u}^0 .
\]
The matrix ${\bf H}^{(3)}$ follows from eqs.\   \eqref{223}, \eqref{225}, \eqref{11} and
\eqref{409}.  

The subsequent general analysis applies only to the 2D lattice for which the vectors ${\bf u}^0$, ${\bf u}^1$, ${\bf u}^2$ are 6-dimensional.  The analogous derivation for the 3D case, which  involves 12-dimensional vectors,  is not considered here, although we note that some explicit low frequency asymptotic  results are given in \S\ref{sec6}. 

\subsection{Effective speeds in 2D lattices}
Consider the equation
\[\lb{403}
 {\bf H}^{(0)}\ma {\bf u}_1 \\ {\bf u}_2 \am={\bf 0}.
\]
Since ${\bf H}_\pm^{(0)}= ({\bf H}_\pm^{(0)})^+>0$ then it is not difficult to show that the
solution of \er{403} satisfies
\[\lb{404}
 {\bf u}_1={\bf u}_2,\ \ {\bf u}_1\perp{\bf e}_b. 
\]
Based on eqs.\ \eqref{403} and \eqref{404}
 we obtain the following result:

The dimension of $\Ker{\bf H}^{(0)}$ \er{403} is equal to $2$ and the basis can
be chosen as
\[\lb{406}
 {\bf u}^{01}=2^{-\frac12}\ma {\bf e}^{01} \\ {\bf e}^{01} \am,\ \
 {\bf u}^{02}=2^{-\frac12}\ma {\bf e}^{02} \\ {\bf e}^{02}\am,\ \ {\bf
 e}^{01}=\ma1 \\ 0 \\ 0\am,\ \ {\bf
 e}^{02}=\ma0 \\ 1 \\ 0\am.
\]
In summary, ${\bf H}^{(0)}$ possesses an eigenvalue $\omega=0$ with multiplicity $2$.  We next obtain the equation that determines   the associated pair of wave speeds.
Using the properties derived previously for  ${\bf H}^{(0)}$ of \eqref{400} it follows that the leading order displacement ${\bf u}^0$ is spanned by $\{{\bf u}^{01}, {\bf u}^{02}\}$, see eq.\ \eqref{406}.

Using the identity
${\bf A}_{ij} + {\bf A}_{ij}^{\perp} = \diag (1,1,0)$ for any pair $ij$, it follows that
\[ \lb{4-4}
(({\bf M}-{\bf
 H}^{(3)}){\bf u}^{0i}\cdot{\bf u}^{0j})_{i,j=1}^2 =
\frac m2 \diag (1,1)
\]
where 
 $m$ is the total mass per unit cell,
\beq{467}
m = m_1+m_2 + \sum\limits_{j\in\mathcal{N}_1}
 \r_{1j} l_{1j}  .
\eeq
The appearance of the total mass is significant, bearing in mind that dynamic lattice models which do not include both flexural and longitudinal waves are known to produce  quasistatic wave speeds with effective mass less than the total mass of the unit cell \cite{Colquitt11}.

Scalar multiplying \er{4-4} by  ${\bf u}^0$
and using ${\bf H}^{(1)}{\bf u}^0\cdot{\bf u}^0=0$
implies
\[\lb{4-3}
\omega_1 = 0,\ \
{\bf u}^1= -  ({\bf H}^{(0)})^{-1} {\bf H}^{(1)}{\bf u}^0
\]
where $({\bf H}^{(0)})^{-1}$ is uniquely defined   acting on the subspace orthogonal to $\spn \{{\bf u}^{01},{\bf u}^{02}\}$.

At O$(k^2)$ we have
\[\lb{413}
 \omega_2({\bf M}-{\bf  H}^{(3)}){\bf u}^0\cdot{\bf u}^{0j} = {\bf H}^{(2)}{\bf u}^0\cdot{\bf
 u}^{0j}+ {\bf H}^{(1)}{\bf u}^1\cdot{\bf u}^{0j} , \ \ j=1,2.
\]
 Hence we
deduce  that the squares of the effective speeds $c_{\rm eff}^2 =\omega_2$ are
eigenvalues of the following $2\ts2$ matrix 
\beq{415}
 {\bf C}_{\rm eff}^2= \frac 2m
\big\{
({\bf H}^{(2)}{\bf u}^{0i}\cdot{\bf u}^{0j})_{i,j=1}^2-
 (({\bf H}^{(0)})^{-1}{\bf H}^{(1)}{\bf u}^{0i}\cdot{\bf H}^{(1)}{\bf
 u}^{0j})_{i,j=1}^2 \big\}
\eeq
where $m$ defined in \er{467} is the total mass per unit cell, ${\bf H}^{(0)}$ is given in \er{400} and
\beq{4-9}
\begin{aligned}
{\bf H}^{(1)} &= \ma {\bf 0} & {\bf A} \\ {\bf A}^+ & {\bf 0} \am ,
\ \
{\bf A} = -i
\sum\limits_{j\in\mathcal{N}_1}
 {\bf P}_{1j}^{(2)} (0) \, ({\bf g}_{j}\cdot \bk),
\\
{\bf H}^{(2)} &= \ma {\bf 0} & {\bf B} \\ {\bf B}^+ & {\bf 0} \am ,
\ \
{\bf B} = \frac 12
\sum\limits_{j\in\mathcal{N}_1}
 {\bf P}_{1j}^{(2)} (0) \, ({\bf g}_{j}\cdot \bk)^2 .
\end{aligned}
\eeq

The expression \er{415} can be simplified as follows, with ${\bf I}_{2,3}=\ma {\bf e}^{01} & {\bf e}^{02}
 \am$, 
\[\lb{effn1}
 {\bf C}_{\rm eff}^2= \frac 1m{\bf I}_{2,3}^{T}({\bf B}+{\bf B}^+-2{\bf A}^+(2{\bf H}_+^{(0)}+{\bf R}_+-2{\bf H}^{(0)}_+{\bf A}_b)^{-1}{\bf A}){\bf
 I}_{2,3} . 
\]
Introducing the matrices
\beq{effn2}
\begin{aligned}
 {\bf B}_1 &=\sum_{j\in\cN_1}\big(
\tilde\mu_{1j}{\bf A}_{1j}  + 12  \lambda_{1j}l_{1j}^{-3}
 {\bf A}_{1j}^{\perp  }\big)({\bf g}_{j}\cdot \bk)^2,
\\
 {\bf B}_2 & = \sum_{j\in\cN_1}\big(
\tilde\mu_{1j}{\bf A}_{1j}  + 6  \lambda_{1j}l_{1j}^{-3} (
 2{\bf A}_{1j}^{\perp  }
+ {\bf A}_{1j}^{b\perp  }) \big)({\bf g}_{j}\cdot \bk),
\\ 
 {\bf B}_3 &=\sum_{j\in\cN_1}\big(
\tilde\mu_{1j}{\bf A}_{1j}  + 3  \lambda_{1j}l_{1j}^{-3} (
 4{\bf A}_{1j}^{\perp  }
+ 2{\bf A}_{1j}^{\perp b }  +  2{\bf A}_{1j}^{b\perp  } +
{\bf A}_b  ) \big),
\end{aligned}
\eeq
we can rewrite \er{effn1} succinctly as
\[\lb{effn5}
 {\bf C}_{\rm eff}^2= \frac 1m{\bf I}^{T}_{2,3}({\bf B}_1
 -{\bf B}_2^{T}{\bf B}_3^{-1}{\bf B}_2){\bf  I}_{2,3}.
\]


\section{2D Examples} \label{sec5}

\subsection{Rectangular lattice}
\begin{figure}[h!]
                             \begin{center}
\setlength{\unitlength}{0.12mm}
\begin{picture}(229,236)(120,-298)
        \put(127,-182){{\ellipse*{15}{15}}} 
        \allinethickness{0.254mm}\path(135,-183)(230,-183) 
        \allinethickness{0.254mm}\path(240,-183)(335,-183) 
        \put(341,-183){{\ellipse*{15}{15}}} 
        \allinethickness{0.254mm}\path(232,-83)(232,-178) 
        \put(231,-75){{\ellipse*{15}{15}}} 
        \put(232,-183){{\ellipse*{15}{15}}} 
        \put(231,-290){{\ellipse*{15}{15}}} 
        \allinethickness{0.254mm}\path(232,-190)(232,-285) 
        \put(250,-160){\makebox(0,0)[cc]{\shortstack{{\Large a$_0$}}}} 
        \put(336,-164){\makebox(0,0)[cc]{\shortstack{{\Large a$_1$}}}} 
        \put(203,-92){\makebox(0,15)[cc]{\shortstack{{\Large a$_2$}}}} 
        \put(122,-170){\makebox(0,15)[cc]{\shortstack{{\Large a$_3$}}}} 
        \put(197,-290){\makebox(0,15)[cc]{\shortstack{{\Large a$_4$}}}} 
\end{picture}
                              \end{center}
\caption{The rectangular lattice cell with mass at ${\bf a}_0$, showing the neighboring masses in the adjoining cells. }
{\label{fig2}}
                           \end{figure}
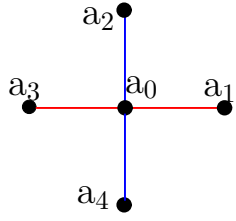

\subsubsection{Dispersion relation}
The unit cell for  the rectangular lattice, shown in Fig.\ \ref{fig2}, possesses a  mass at the central node.   Enforcing the equilibrium condition at the single mass and the Bloch-Floquet condition, it may be shown that the equations of motion for this structure reduce to 
\begin{equation} \label{589213}
\sum\limits_{j=1,2,3,4}\big( {\bf P}_{0j}^{(1)} - {\bf P}_{0j}^{(2)}{\bf e}^{i{\bf k}\cdot{\bf g}_j} \big){\bf u}_0 = \omega^2 {\bf M}_0 {\bf u}_0,
\ \ {\bf M}_0 = \diag(m_0,m_0,I_0).
\end{equation}
The derivation is entirely similar to that for the hexagonal lattice in Sections \ref{sec2} and \ref{sec3}, with the same notation employed.

We assume the members are of two types: $1$ for horizontal, and $2$ for vertical members,
with parameters denoted  by  $\rho_j, {\bf K}^{(j)}$, etc.\ $j=1,2$. Then it may be shown that eq.\
 \eqref{589213} becomes 
\beq{-5}
\begin{pmatrix}
\begin{matrix}
\tilde{\mu}_1 \tilde{s}_1(\cot\tilde{s}_1 -\csc\tilde{s}_1 \cos \tilde{k}_x)
\\
+
\l_2 (K^{(2)}_{11}+K^{(2)}_{13}\cos \tilde{k}_y)
\end{matrix}
& 0 & i\l_2K^{(2)}_{14}\sin \tilde{k}_y
\\ 0 &
\begin{matrix}
\tilde{\mu}_2 \tilde{s}_2(\cot\tilde{s}_2 -\csc\tilde{s}_2  \cos \tilde{k}_y )
\\
+\l_1 (K^{(1)}_{11}+K^{(1)}_{13}\cos \tilde{k}_x)
\end{matrix}
& -i\l_1 K^{(1)}_{14}\sin \tilde{k}_x
\\ -i\l_2 K^{(2)}_{14}\sin \tilde{k}_y  & i\l_1 K^{(1)}_{14}\sin \tilde{k}_x  &
 \begin{matrix}
\l_1 (K^{(1)}_{22} + K^{(1)}_{24} \cos \tilde{k}_x )
\\
+\l_2( K^{(2)}_{22} + K^{(2)}_{24} \cos \tilde{k}_y )
\end{matrix}
\end{pmatrix}
 {\bf u}_0
=  \frac {\omega^2}2 {\bf M}_0{\bf u}_0
\eeq
where ${\bf k}= (k_x,k_y)$ and
$
\tilde{k}_x = l_1k_x$, $\tilde{k}_y=l_2k_y$.


\subsubsection{Quasi-static effective speeds for the rectangular lattice}

Using ${\bf k}=k\bk$ the second order asymptotics of \er{-5} \rev2{are}
\[\lb{eff1}
 (k^2{\bf A}+\omega^2{\bf B}+k{\bf D}+{\bf E})({\bf u}_0+k{\bf u}_1+k^2{\bf
 u}_2)={\bf 0}
\]
with matrices of the form
\[\lb{eff2}
 {\bf A}=\diag(A_j),\ \ {\bf B}=\diag(B_j),\ \ {\bf
 E}=\diag(0,0,E),\ \ {\bf D}=\ma
                               0 & 0 & d_1 \\
                               0 & 0 & d_2 \\
                               d_1^* & d_2^* & 0
                             \am
\]
where (in the following calculations we do not need exact values of
$A_3$, $B_3$)
\[\lb{eff2a}
\begin{aligned}
 A_1&=\frac 12{\m_1 l_1 \k_x^2}+ {6\l_2l_2^{-1}\k_y^2},\ \
 A_2=\frac 12{\m_2 l_2\k_y^2 }+ {6\l_1 l_1^{-1}\k_x^2},
\ \
 E=6\l_1 l_1^{-1}+6\l_2 l_2^{-1} ,
\\
B_1&=B_2=- \frac 12(\r_1 l_1+\r_2 l_2 +m_0),
\ \ d_1={6i\l_2l_2^{-1} \k_y},\ \ d_2=-{6i\l_1 l_1^{-1}\k_x} .
\end{aligned}
\]

Substituting $\omega=c_jk$, ${\bf u}_i={\bf u}_{ji}$, $j=1,2$ (because
for $\omega,k=0$ we have two solutions) into \er{eff1} we obtain
\begin{eqnarray} \lb{eff5}
 k^0&:&{\bf E}{\bf u}_{j0}={\bf 0}, \lb{eff3} \notag \\
 k^1&:&{\bf D}{\bf u}_{j0}+{\bf E}{\bf u}_{j1}={\bf 0}, \lb{eff4}  \\
 k^2&:&({\bf A}+c_j^2{\bf B}){\bf u}_{j0}+{\bf D}{\bf u}_{j1}+{\bf E}{\bf
 u}_{j2}={\bf 0}.\notag
\end{eqnarray}
Scalar multiplying the O$(k^2)$ equation by  ${\bf u}_{j0}$ and using
\er{eff2}-\er{eff4} with self-adjointness of all matrices we deduce
that
\[\lb{eff6}
 \ma{\bf u}_{10} & {\bf u}_{20}\am^{T}({\bf A}+c_j^2{\bf B}){\bf u}_{j0}-\frac1E\ma{\bf u}_{10} & {\bf u}_{20}\am^{T}
 \ma d_1 \\ d_2 \\ 0 \am\ma d_1^* & d_2^* & 0 \am{\bf
 u}_{j0}={\bf 0}.
\]
Using \er{eff2a} we can rewrite the effective
equations \er{eff6} as 
\[\lb{eff9}
 \ma
   \m_1l_1\k_x^2+\frac{12\k_y^2}{ \l_1^{-1}l_1+\l_2^{-1}l_2}
    & -\frac{12\k_x\k_y}{\l_1^{-1}l_1 +\l_2^{-1}l_2} \\
   -\frac{12\k_x\k_y}{\l_1^{-1}l_1 +\l_2^{-1}l_2}
    & \m_2l_2\k_y^2+\frac{12\k_x^2}{\l_1^{-1}l_1 +\l_2^{-1}l_2}
 \am{\bf v}_{j0}=mc_j^2{\bf v}_{j0}
\]
with, as expected \cite{Colquitt11}, the total mass per unit cell 
\[\lb{eff10}
 m=m_0 + \r_1 l_1 +\r_2 l_2 .
\]
The equation \er{eff9} with constant matrix has two solutions:
effective speeds $c_j^2$ and corresponding constant displacements
${\bf v}_{j0}$, $j=1,2$, which are eigenvalues and eigenvectors of
the left matrix divided by $m$.

\subsubsection{Numerical example}

We consider wave propagation in the $x-$direction $(\tilde{k}_y=0)$
    in which case the solutions of \eqref{-5} simplify as follows:  (i) a
 quasi-longitudinal solution ${\bf u}_0 = (1,0,0)^T$ with 
 $\tilde{k}_x$  given explicitly in terms of $\omega$ from
\beq{46-}
\cos \tilde{k}_x  = \cos\tilde{s}_1 +
\Big(
{\l_2 }{}(K^{(2)}_{11}+K^{(2)}_{13}) - \frac 12 {m_0 \omega^2}
\Big)\, \frac{ \sin\tilde{s}_1}{ \tilde{\mu}_1\tilde{s}_1} .
\eeq
Note that this mode couples longitudinal effects in the $x-$direction with flexural effects in the  $y-$direction.
(ii) a quasi-flexural solution ${\bf u}_0 = (0,a,b)^T$ with  dispersion relation in the form of a quadratic equation for $\cos \tilde{k}_x $
\bal{-45}
&\Big(\l_1 (K^{(1)}_{11}+K^{(1)}_{13}\cos \tilde{k}_x )
+\tilde{\mu}_2\tilde{s}_2 (\cot\tilde{s}_2-\csc\tilde{s}_2 ) - \frac{1}{2}m_0\omega^2\Big)
\notag \\
& \quad \times
\Big( \l_1 (K^{(1)}_{22} + K^{(1)}_{24} \cos \tilde{k}_x )
+\l_2( K^{(2)}_{22} + K^{(2)}_{24}  )
-\frac{1}{2}I_0\omega^2\Big)
-\big(\l_1 K^{(1)}_{14}\sin \tilde{k}_x \big)^2=0.
\eal
Solutions of this dispersion relation   couple the flexural wave in the $x-$direction with both the longitudinal and flexural waves in the $y-$direction.

We consider a lattice with square unit cell  of size $L^2$, with all members the same and of thickness $t$ (and therefore radius of gyration $\k = t/\sqrt{12}$).  The dimensions and properties used are given  in Table \ref{tab1}, which corresponds to an example considered in   \cite{Leamy12}.
\begin{table}[ht]
\caption{Parameters of the square lattice.} 
\centering
\begin{tabular}{ |c | c | c | c| c |}
  \hline
  $E$ (GPa) & $\nu$ & $\rho^V$ (kg/m$^3$)& $L$ (mm)& $t$ (mm) \\
    \hline  \hline
  70 & .33 & 2.7 $\cdot$ 10$^3$ & 10 & 1 \\
  \hline
\end{tabular}
\label{tab1}
\end{table}
Results based on eq.\ \eqref{-5} are shown in Fig.\  \ref{fig1numeric} along with a comparison against results found using FEM (COMSOL). 
The  two types of wave solutions defined by eqs.\ \er{46-} and \er{-45}  are distinguished in Fig.\  \ref{fig1numeric}.
Based on the comparison with the FEM calculations in Fig.\  \ref{fig1numeric} it is evident  that the present theory provides an excellent match to the first six Floquet branches for waves propagating in the $x-$direction.

 \begin{figure}[h!]  
\begin{center}
\subfloat[]{
 \includegraphics[width=2.5in]{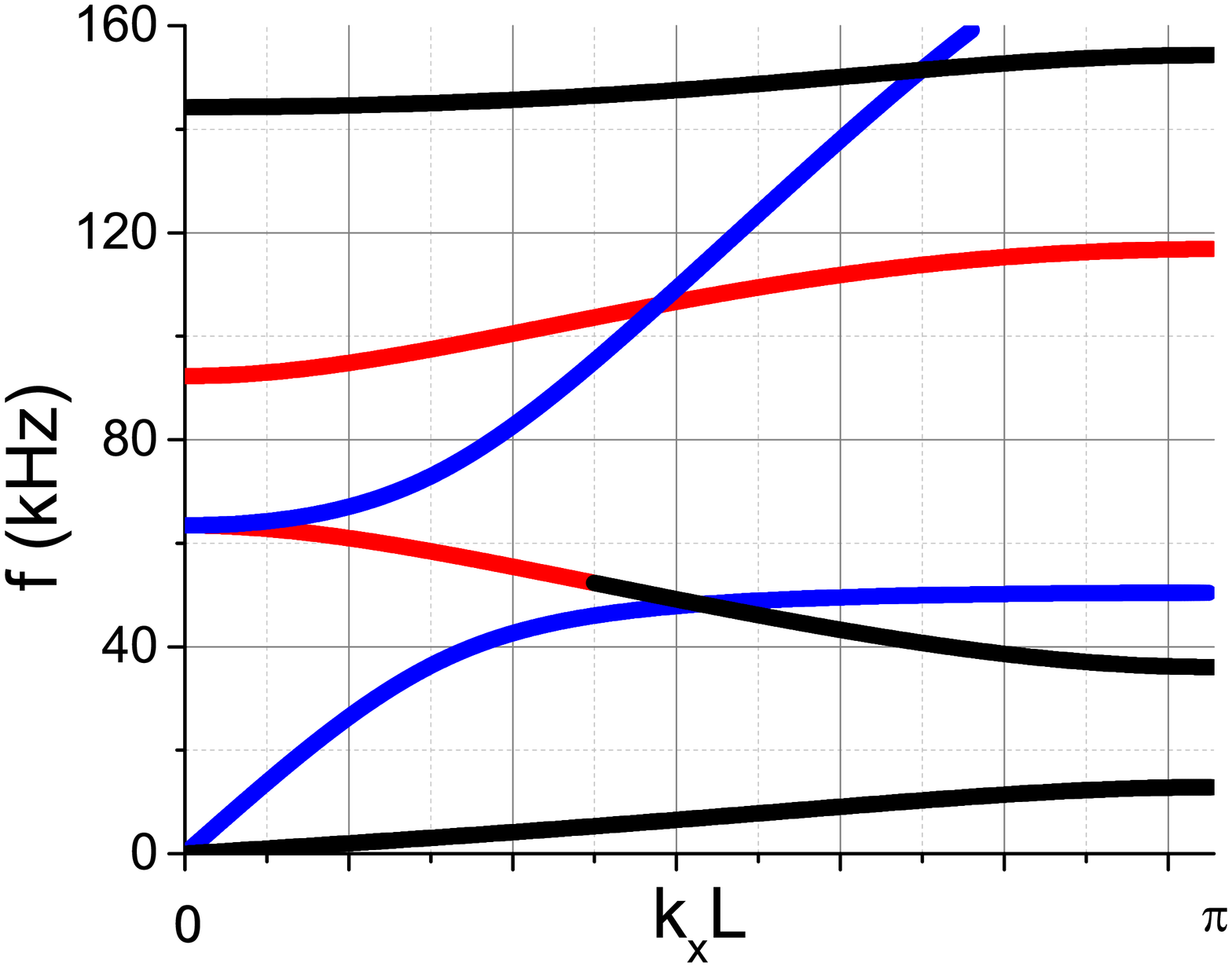}
}
\subfloat[]{
 \includegraphics[width=2.5in]{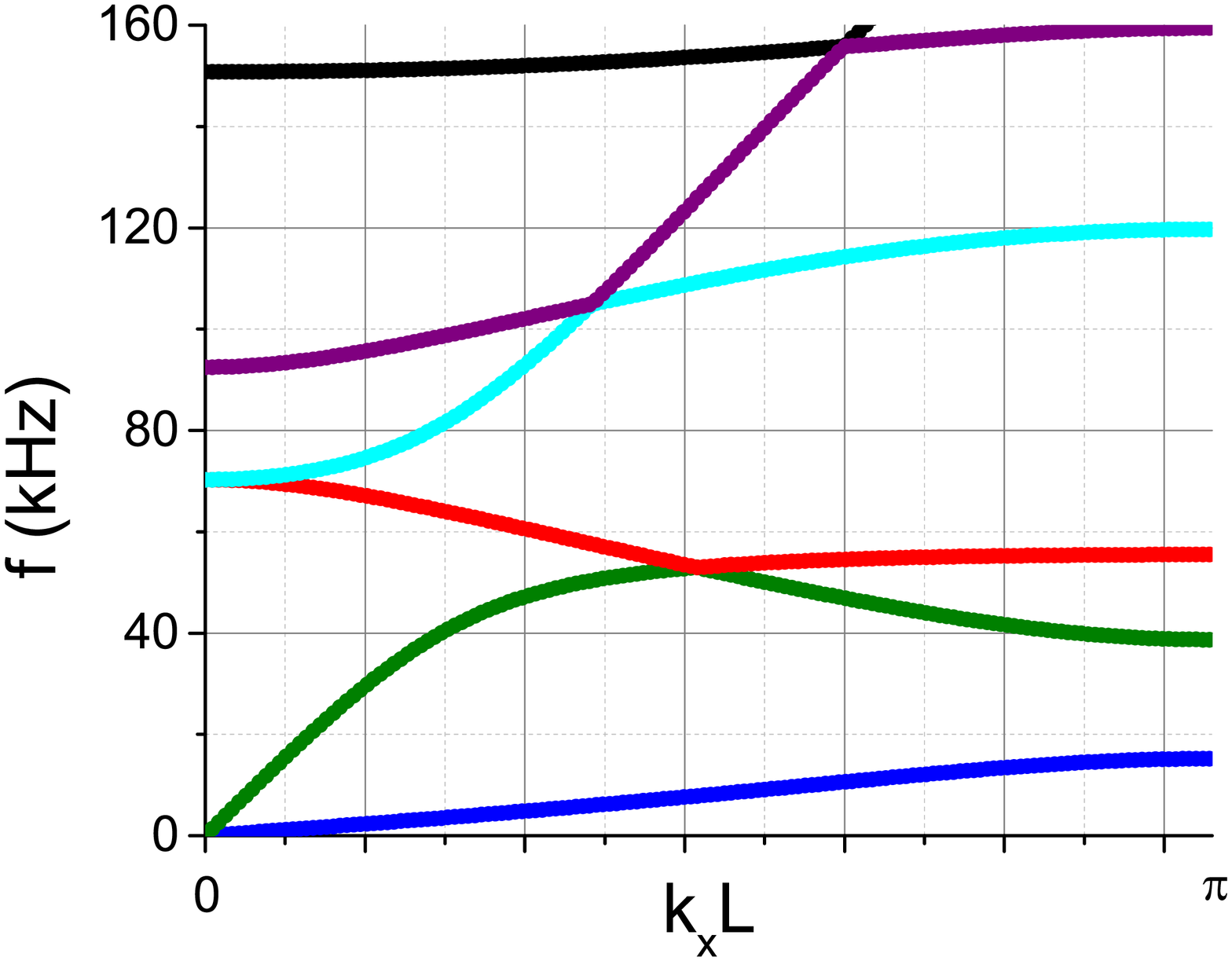}
}
\end{center}
 \caption{Dispersion curves of the square lattice of Table \ref{tab1} for $k_y=0$.  (a)  The blue curves correspond to  quasi-longitudinal motion described by eq.\ \eqref{46-}; the black and red curves correspond to the pair of quasi-transverse solutions      described by
eq.\ \eqref{-45}.  (b)  Dispersion curves calculated using FEM (COMSOL). 
}
\label{fig1numeric}
\end{figure}

\subsection{Hexagonal lattice}

\begin{figure}[h!] \label{hex_unit}
\centering
\subfloat[]{
\setlength{\unitlength}{0.15mm}
\begin{picture}(201,175)(160,-260)
        \put(331,-179){\makebox(0,0)[cc]{\shortstack{{\Large a$_1$}}}} 
        \put(193,-117){\makebox(0,15)[cc]{\shortstack{{\Large a$_5$}}}} 
        \put(227,-175){\makebox(0,15)[cc]{\shortstack{{\Large a$_2$}}}} 
        \put(192,-260){\makebox(0,15)[cc]{\shortstack{{\Large a$_6$}}}} 
        \put(332,-260){\makebox(0,15)[cc]{\shortstack{{\Large a$_4$}}}} 
        \put(332,-115){\makebox(0,15)[cc]{\shortstack{{\Large a$_3$}}}} 
        \put(214,-181){{\ellipse*{15}{15}}} 
        \allinethickness{0.254mm}\path(215,-180)(305,-180) 
        \put(305,-180){{\ellipse*{15}{15}}} 
        \put(353,-107){{\ellipse*{15}{15}}} 
        \put(167,-107){{\ellipse*{15}{15}}} 
        \put(352,-252){{\ellipse*{15}{15}}} 
        \allinethickness{0.254mm}\path(305,-180)(355,-105) 
        \allinethickness{0.254mm}\path(305,-180)(355,-255) 
        \allinethickness{0.254mm}\path(165,-105)(215,-180) 
        \allinethickness{0.254mm}\path(165,-255)(215,-180) 
        \put(167,-252){{\ellipse*{15}{15}}} 
\end{picture}
}
\hspace{20pt}
\subfloat[]{
 \includegraphics[width=0.3\textwidth]{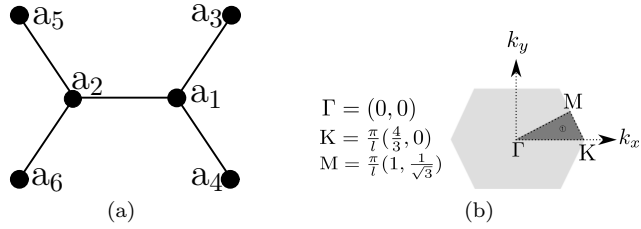} \label{ibz_hex}
}
  \caption{Hexagonal lattice. (a) The unit cell. (b) The irreducible Brillouin zone \cite{Maldovanbook09}.}
\end{figure}

\subsubsection{Quasi-static effective speeds for the hexagonal lattice}

Consider the special case in which  the lattice is a regular hexagon with uniform properties $\mu$, $\lambda$ and $l$.   It follows from eq.\ \eqref{effn5} that 
\beq{-31}
{\bf C}_{\rm eff}^2  = 
\frac{3l}{2m} \,\Big[ \Big(\frac 1\mu + \frac {l^2}{12\lambda}
\Big)^{-1} \diag (1,1) + \frac\mu 2 \bk\bk^T
\Big] 
\eeq   
The eigenvectors of the matrix  ${\bf C}_{\rm eff}^2$ are then  purely longitudinal and transverse, i.e. parallel and perpendicular to $\bk$, with wave speeds $c_L$ and $c_T$,  respectively, where
\beq{-32}
c_T^2 = \frac{3l}{2m} \, \Big(\frac 1\mu + \frac {l^2}{12\lambda}
\Big)^{-1},
\quad
c_L^2 = c_T^2 + \frac{3l}{4m} \mu .
\eeq   

\subsubsection{Numerical result}
We consider an example for which all members have the same uniform properties and are arranged in a regular hexagonal lattice.  
The numeric computations are based on the  properties in Table \ref{tab2} and the path of the wave vector taken is along the perimeter of the Brillouin zone shown in Fig.\ \ref{ibz_hex}.

\begin{table}[h!]
\caption{Hexagonal lattice parameters.} 
\centering
\begin{tabular}{| c | c | c | c| c |}
  \hline
  $E$ (GPa) & $\nu$ & $\rho^V$ (kg/m$^3$)& $l$ (mm)& $t$ (mm) \\
    \hline  \hline
  70 & .33 & 2.7 $\cdot$ 10$^3$ & 10 &  1 \\
  \hline
\end{tabular}
\label{tab2}
\end{table}
The dispersion curves in Fig.\ \ref{figsec2hex2numeric}(a)  \reva{were obtained from eq.\ \eqref{226} } using a combination of minimum value threshold and minimum peak finding methods for the $6\ts 6$  determinant evaluated on  a discretized grid of  wave vector and frequency.  This provides a fast solution technique, which can be refined by taking smaller grid steps.   Figure \ref{figsec2hex2numeric} shows that the  dispersion curves  computed by the present simplified theory  agree well with those found using FEM.  A close comparison shows some small deviations from the FEM results (which can safely be considered as an accurate benchmark) but the overall agreement is remarkable considering the simplicity of the present approach.   \rev2{The hexagonal system displays a strong one-wave effect between approximately 15 and 30 kHz. In this range the dispersion is weak, as indicated by the almost straight line  branches.  Furthermore, the hexagonal symmetry ensures isotropy in the long-wavelength limit, which is the original reason \cite{Norris11mw} for our interest in this particular structure. 
}

\reva{Note that the  roots obtained in Fig.\ \ref{figsec2hex2numeric}(a) were numerically checked using a symbolic algebra -generated expression for the determinant of eq.\ \eqref{226}.  Although significant speedup in computing time was not observed, this was not the primary purpose and future work could use such very lengthy but precise expressions to better computational advantage.
}

 \begin{figure}[h!]  
\begin{center}
\subfloat[]{
 \includegraphics[width=2.5in]{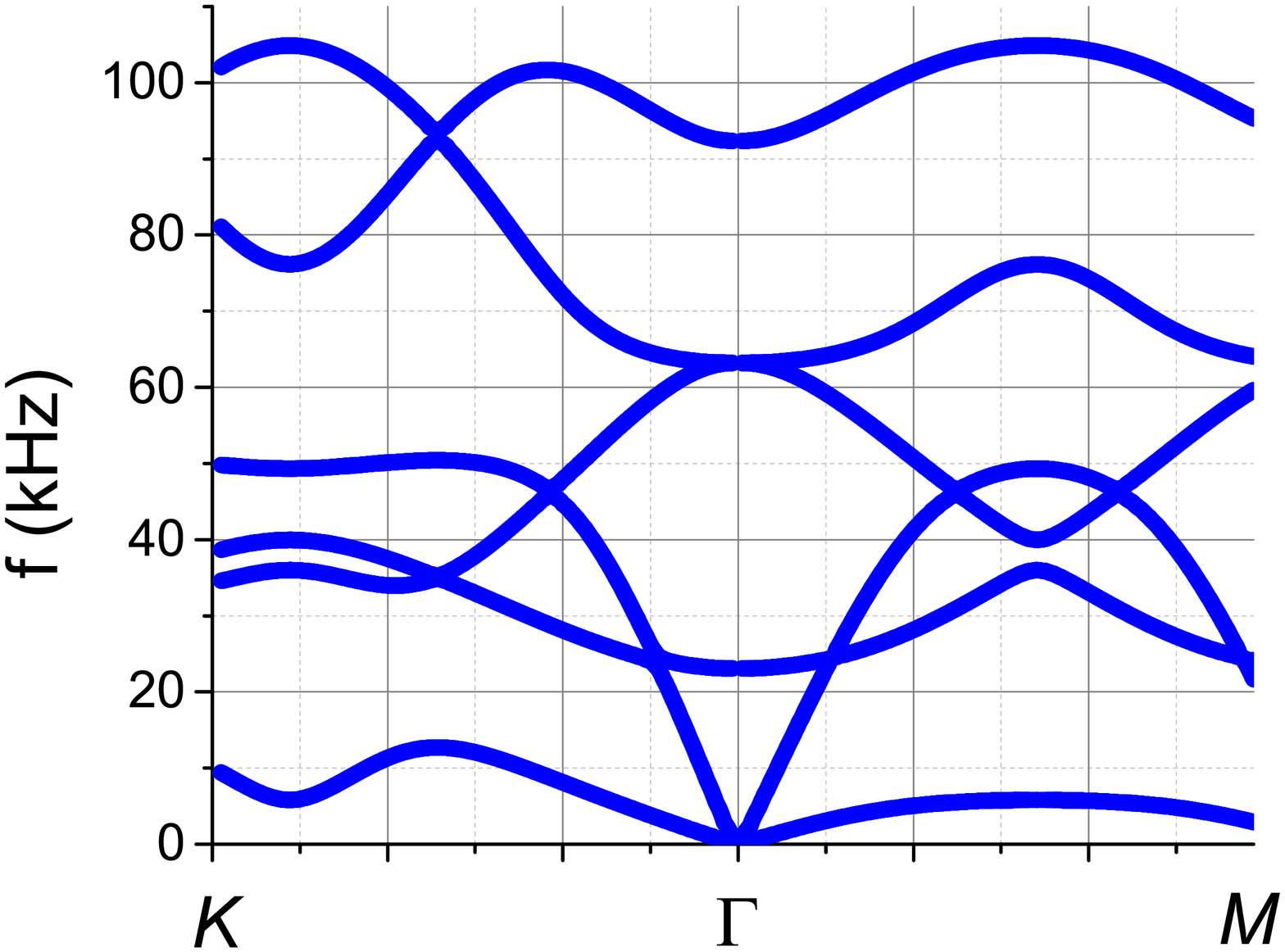}
}
\subfloat[]{
 \includegraphics[width=2.5in]{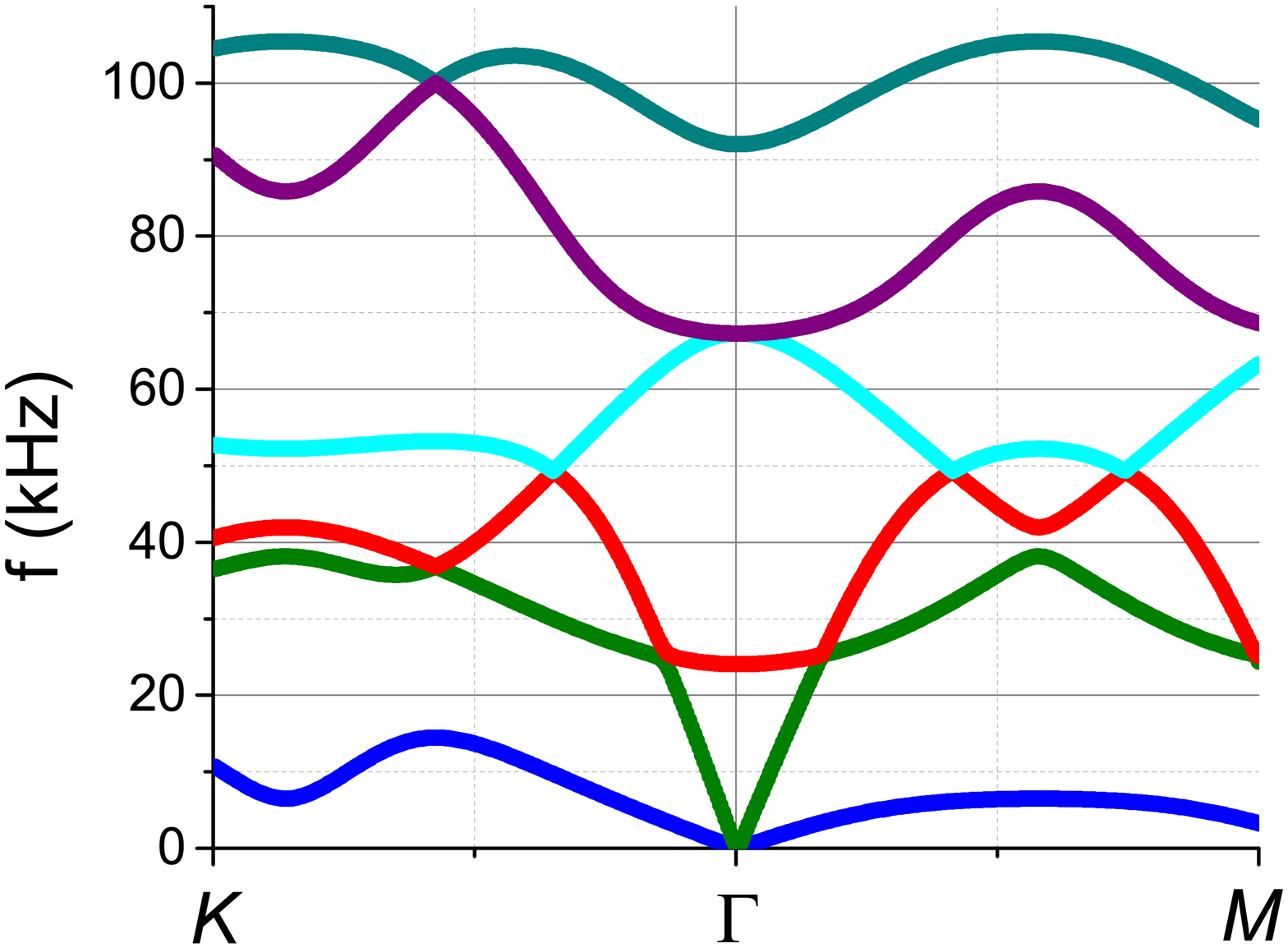}
}
\end{center}
\caption{Dispersion curves of the regular hexagonal  lattice with properties in Table \ref{tab2}.  (a)  The first six Floquet branches for wave-vector along the perimeter of the Brillouin zone.  (b)  Dispersion curves calculated using FEM (COMSOL).
\rev2{Note the almost non-dispersive one-wave behaviour between 15 and 30 kHz.}}
 \label{figsec2hex2numeric}
\end{figure}

\section{3D Examples}  \lb{sec6}

\subsection{Cubic lattice}
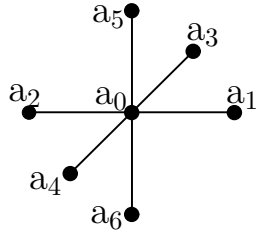
\begin{figure}[H] \label{cubic_cell_lattice}
  \centering
\setlength{\unitlength}{0.18mm}
\begin{picture}(161,173)(180,-260)
        \put(245,-170){\makebox(0,0)[cc]{\shortstack{{\Large a$_0$}}}} 
        \put(341,-169){\makebox(0,0)[cc]{\shortstack{{\Large a$_1$}}}} 
        \put(243,-117){\makebox(0,15)[cc]{\shortstack{{\Large a$_5$}}}} 
        \put(182,-175){\makebox(0,15)[cc]{\shortstack{{\Large a$_2$}}}} 
        \put(242,-265){\makebox(0,15)[cc]{\shortstack{{\Large a$_6$}}}} 
        \put(197,-240){\makebox(0,15)[cc]{\shortstack{{\Large a$_4$}}}} 
        \put(312,-130){\makebox(0,15)[cc]{\shortstack{{\Large a$_3$}}}} 
        \put(185,-180){{\ellipse*{10}{10}}} 
        \allinethickness{0.254mm}\path(185,-180)(335,-180) 
        \put(260,-180){{\ellipse*{10}{10}}} 
        \put(335,-180){{\ellipse*{10}{10}}} 
        \put(260,-105){{\ellipse*{10}{10}}} 
        \put(260,-255){{\ellipse*{10}{10}}} 
        \allinethickness{0.254mm}\path(260,-105)(260,-255) 
        \put(305,-135){{\ellipse*{10}{10}}} 
        \put(215,-225){{\ellipse*{10}{10}}} 
        \allinethickness{0.254mm}\path(305,-135)(215,-225) 
\end{picture}

  \caption{Cubic unit cell}
\end{figure}

\subsubsection{Dispersion relations}
Similar to the rectangular lattice, the equation of motion can be written as
\begin{equation} \label{cubic_eom}
\sum\limits_{j=1,2,3,4,5,6}\big( {\bf P}_{0j}^{(1)} - {\bf P}_{0j}^{(2)}{\bf e}^{i{\bf k}\cdot{\bf g}_j} \big){\bf u}_0 = \omega^2 {\bf M}_0 {\bf u}_0,
\ \ {\bf M}_0 = \diag(m_0,m_0,m_0,I_0,I_0,I_0).
\end{equation}

We assume the members are of three types: 1, 2, 3  for the $x$, $y$, and  $z$-directions, respectively, with parameters denoted by $\rho_j$, ${\bf K}^{(j)}$, etc. $j=1,2,3$, then eq.\
 \eqref{cubic_eom} becomes,
\begin{footnotesize}
\begin{align}\label{-4}
&\begin{pmatrix} \zeta_1 & 0&0&0 & i\l_3K^{(3)}_{14}\sin \tilde{k}_z & -i\l_2K^{(2)}_{14}\sin \tilde{k}_y\\ 0& \zeta_2 & 0 & -i\l_3K^{(3)}_{14}\sin \tilde{k}_z & 0 & i\l_1K^{(1)}_{14}\sin \tilde{k}_x\\ 0&0& \zeta_3& i\l_2K^{(2)}_{14}\sin \tilde{k}_y & -i\l_1K^{(1)}_{14}\sin \tilde{k}_x &0\\ 0& i\l_3K^{(3)}_{14}\sin \tilde{k}_z & -i\l_2K^{(2)}_{14}\sin \tilde{k}_y & \zeta_4&0&0\\ -i\l_3K^{(3)}_{14}\sin \tilde{k}_z & 0 & i\l_1K^{(1)}_{14}\sin \tilde{k}_x & 0 &  \zeta_5 & 0\\i\l_2K^{(2)}_{14}\sin \tilde{k}_y & -i\l_1K^{(1)}_{14}\sin \tilde{k}_x & 0 & 0 & 0 & \zeta_6\end{pmatrix}   2 {\bf u}_0
\notag \\
& \qquad \qquad 
=\omega^2{\bf M}_0{\bf u}_0 ,
\end{align}
\end{footnotesize}
where ${\bf k}= (k_x,k_y,k_z)$, 
$( \tilde{k}_x ,  \tilde{k}_y,  \tilde{k}_z) = (l_1k_x,l_2k_y,l_3k_z)$ and 
\begin{equation}
\begin{aligned}
&\zeta_1=\l_2 (K^{(2)}_{11}+K^{(2)}_{13}\cos \tilde{k}_y)+\l_3 (K^{(3)}_{11}+K^{(3)}_{13}\cos \tilde{k}_z)
+\tilde{\mu}_1 \tilde{s}_1(\cot\tilde{s}_1 -\csc\tilde{s}_1 \cos \tilde{k}_x),
\\&\zeta_2=\l_3 (K^{(3)}_{11}+K^{(3)}_{13}\cos \tilde{k}_z)+\l_1 (K^{(1)}_{11}+K^{(1)}_{13}\cos \tilde{k}_x)
+\tilde{\mu}_2 \tilde{s}_2(\cot\tilde{s}_2 -\csc\tilde{s}_2  \cos \tilde{k}_y),
\\&\zeta_3=\l_1 (K^{(1)}_{11}+K^{(1)}_{13}\cos \tilde{k}_x)+\l_2(K^{(2)}_{11}+K^{(2)}_{13}\cos \tilde{k}_y )
+\tilde{\mu}_3 \tilde{s}_3(\cot\tilde{s}_3 -\csc\tilde{s}_3  \cos \tilde{k}_z ),
\\&\zeta_4=\l_2 (K^{(2)}_{22}+K^{(2)}_{24}\cos \tilde{k}_y)+\l_3 (K^{(3)}_{22}+K^{(3)}_{24}\cos \tilde{k}_z),
\\&\zeta_5=\l_3 (K^{(3)}_{22}+K^{(3)}_{24}\cos \tilde{k}_z)+\l_1 (K^{(1)}_{22}+K^{(1)}_{24}\cos \tilde{k}_x),
\\&\zeta_6= \l_1(K^{(1)}_{22}+K^{(1)}_{24}\cos \tilde{k}_x)+\l_2 (K^{(2)}_{22}+K^{(2)}_{24}\cos \tilde{k}_y).
\end{aligned}
\end{equation}

\subsubsection{Quasi-static effective elastic moduli for the cubic lattice}
Considering wave propagation in the $(100)$ and $(110)$ directions of the lattice with pure cubic symmetry ($l_1=l_2=l_3$ etc.) and taking the low frequency limit, we obtain
\beq{eff_moduli}
C_{11}= {EA}/{l^2},\ \ C_{66}={6EI}/{l^4},\ \ C_{12}=0, 
\ \  \rho_{\text {eff}} = (3\rho Al+3m_0)/l^3 . 
\eeq
The moduli are in agreement with known results, e.g. \cite{Norris14}, and the effective mass density is, as expected, identical to the actual density. 
$C_{12}=0$ indicates that Poisson's ratio $\nu_{12}=0$ which can be interpreted as applying a displacement in the $(100)$ direction does not cause deformation in the $(010)$ direction.

\subsubsection{Example: wave propagation in the $x-$direction}
We consider wave propagation along one axis of a lattice structure with uniform material  and structural properties as given  in Table 3 and with members of square cross-section.   
\begin{table}[h!]
\caption{Cubic lattice parameters.} 
\centering
\begin{tabular}{| c | c | c | c| c |}
  \hline
  $E$ (GPa) & $\nu$ & $\rho^V$ (kg/m$^3$)& $l$ (mm)& $t$ (mm) \\
    \hline  \hline
  70 & .33 & 2.7 $\cdot$ 10$^3$ & 10 &  1 \\
  \hline
\end{tabular}
\label{tab3}
\end{table}
Setting  $\tilde{k}_y=\tilde{k}_z=0$, we find that  the first pure-longitudinal solution ${\bf u}_0 = (1,0,0,0,0,0)^T$ of \eqref{-4},  has wavenumber  $\tilde{k}_x$ in terms of $\omega$ as
\begin{equation}\label{426000}
\cos \tilde{k}_x  = \cos\tilde{s} + 
\Big(2
\l (K_{11}+K_{13}) - \frac 12 m_0 \omega^2
\Big) \frac{\  \sin\tilde{s}}{ \tilde{\mu} \tilde{s}} .
\end{equation}
The flexural solution ${\bf u}_0 = (0,1,l,0,\alpha,\beta)^T$ reduces the $6\times6$ equation of motion matrix to a $4\times4$ one. 
\begin{equation}
\begin{pmatrix}B -  m_0\omega^2&0&0&D\\0&B -  m_0\omega^2&-D&0\\0&D&C - I_0\omega^2&0\\-D&0&0&C -  I_0\omega^2 \end{pmatrix}\begin{pmatrix}1\\l\\ \alpha\\ \beta \end{pmatrix} = {\bf 0}, 
\end{equation}
where
\begin{equation}
\begin{aligned}
&B=2\tilde{\mu} \tilde{s}(\cot\tilde{s}-\csc\tilde{s})+2\l \big( 2K_{11}+K_{13}(\cos\tilde{k}_x+1)\big) ,
\\&C=2\l\big( 2K_{22}+K_{24}(\cos\tilde{k}_x+1)\big) ,
\\&D=i2\l K_{14}\sin\tilde{k}_x.
\end{aligned}
\end{equation}
Then calculate the determinant to obtain  the flexural dispersion relation
\begin{equation}\label{430000}
\begin{aligned}
&\Big(\l\big(2K_{11}+K_{13}(\cos \tilde{k}_x+1)\big)+\tilde{\mu}\tilde{s}(\cot\tilde{s}-\csc\tilde{s}) - \frac{1}{2}m_0\omega^2\Big) 
\\
& \quad \times
\Big( \l\big(2K_{22} + K_{24} (\cos \tilde{k}_x +1 )\big)
-\frac{1}{2}I_0\omega^2\Big)
-\big(\l K_{14}\sin \tilde{k}_x \big)^2=0.
\end{aligned}
\end{equation}
\rev2{
In addition to the propagating wave branches   the model also displays pure resonances.  These are modes that are independent of $k_x$ and hence  non-propagating, i.e. with zero group velocity.  They correspond to the generalized displacement
 ${\bf u}_0 = (0,0,0,1,0,0)^T$ which 
 represents flexural resonances (deflection in $x$-direction) of the beams oriented in the $y$- and $z$-directions.    The mode is a solution of  eq.\ \eqref{-4} at resonance frequencies 
that satisfy 
\beq{-56-}
2\l \big(K_{22}+K_{24}\big) - \omega^2 I_0 =0. 
\eeq
In the case considered with $I_0=0$, eq.\ \eqref{-56-}  reduces to 
\beq{-5=7}
\Big( \sin \frac{\gamma l}2 \cosh \frac{\gamma l}2 + \cos \frac{\gamma l}2 \sinh \frac{\gamma l}2 \Big)
\, \sin \frac{\gamma l}2  = 0
\eeq
where $\gamma$ is the flexural wavenumber of Euler beam theory.  The first two lowest solutions of eq.\ \eqref{-5=7} are $\gamma l = 1.5000 \pi$ and $2\pi$. 
}

The dispersion curves for the cubic lattice  are shown in Fig. \ref{fig:dispersion}. \rev2{The analytic results for the propagating wave branches eqs.\ \eqref{426000} and \eqref{430000} match with the FEM simulation.  The first two resonance frequencies of eq.\ \eqref{-5=7} are at 
$51.949$ kHz and $92.354$ kHz, and are shown as  flat branches in Fig.\ \ref{fig:dispersion}(a).  
The first/lowest solution corresponds to the flat branch in Fig.\ \ref{fig:dispersion}(b).  The branch in Fig.\ \ref{fig:dispersion}(b) corresponding to the $92.354$ kHz resonance  shows  slight variation with wavenumber, but is well approximated by the flat branch in Fig.\ \ref{fig:dispersion}(a).
We can conclude from the comparison in Fig.\ \ref{fig:dispersion} that the analytical model predicts the first eight  branches to a remarkable degree of approximation. 
}

\begin{figure}[h!]
     \begin{center}
     \subfloat[]{
         \includegraphics[width=2.5in]{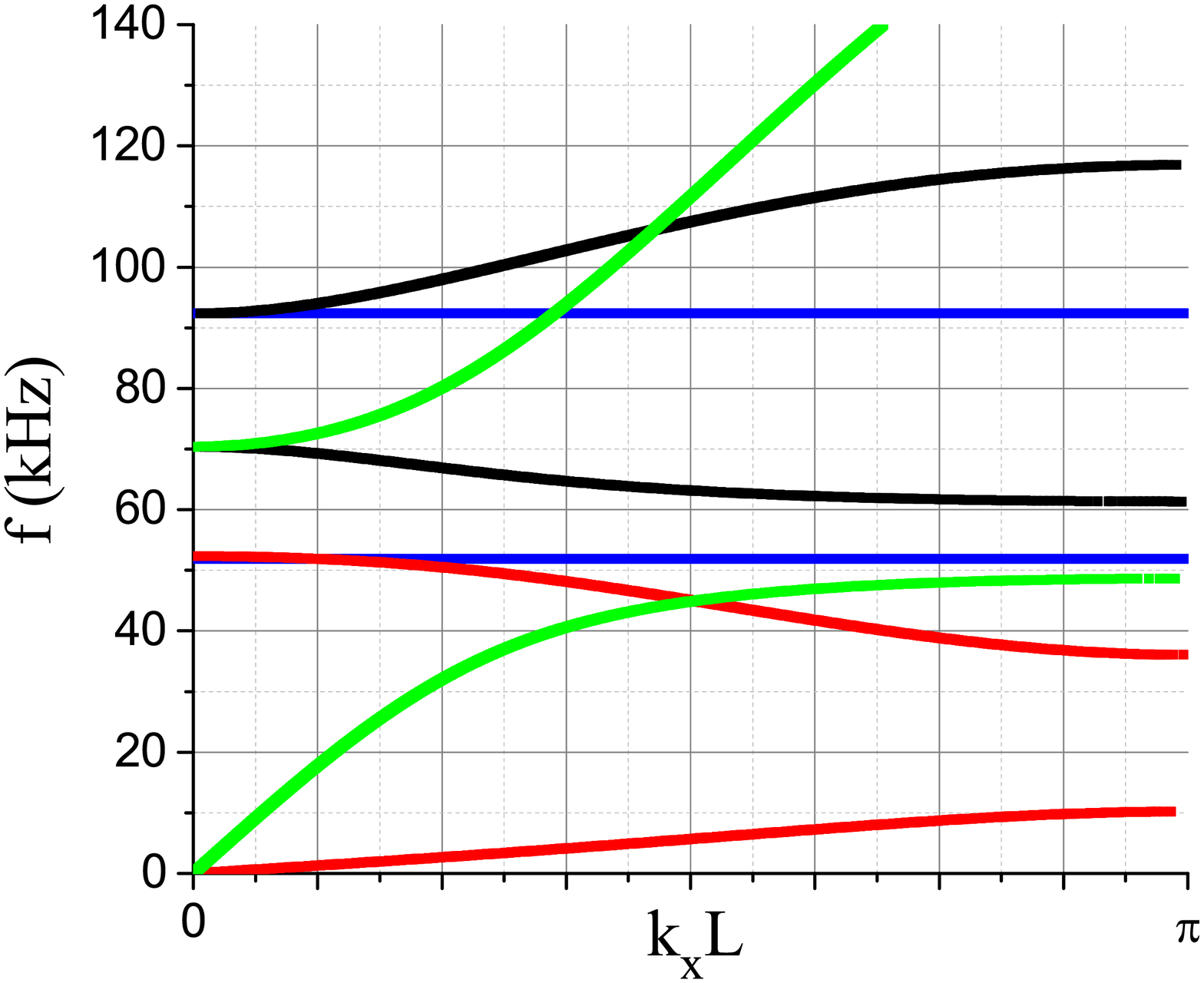}
         }
     \subfloat[]{
         \includegraphics[width=2.5in]{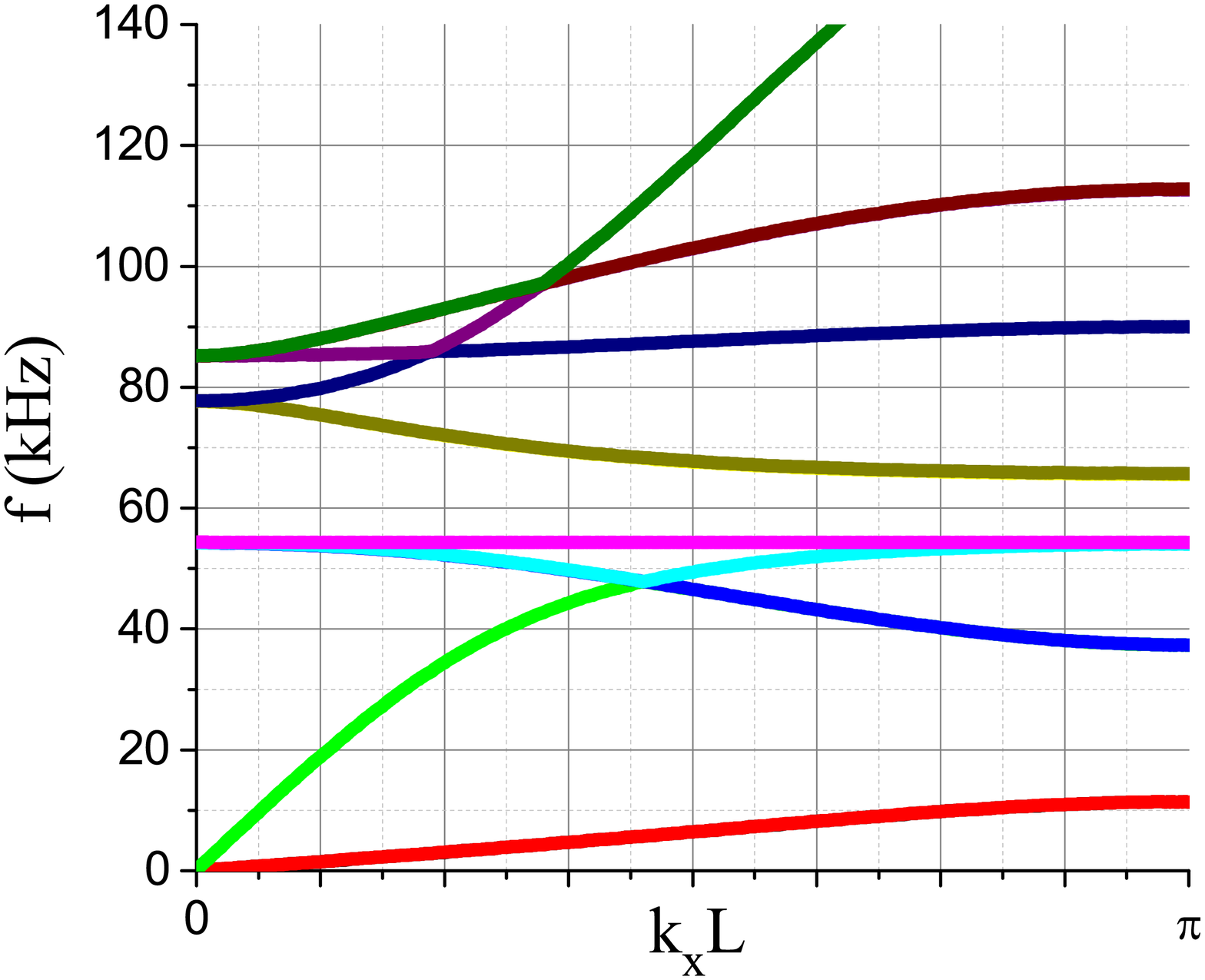}
         }
     \end{center}
     \caption{Dispersion curves of the cubic lattice of Table \ref{tab3}. (a) The green lines are dispersion curves of longitudinal waves, the black and red lines are dispersion curves of shear waves, and the blue curves are flexural resonances of the beams oriented in the $y$- and $z$-directions.  (b) Dispersion curves calculated using FEM (COMSOL). }\label{fig:dispersion}
\end{figure}

\subsection{Tetrahedral lattice}

\begin{figure}[H]
\centering
\subfloat[]{
\setlength{\unitlength}{0.18mm}
\begin{picture}(228,132)(160,-245)
        \put(230,-180){{\ellipse*{10}{10}}} 
        \allinethickness{0.254mm}\path(230,-180)(310,-180) 
        \put(310,-180){{\ellipse*{10}{10}}} 
        \allinethickness{0.254mm}\path(310,-180)(365,-125) 
        \allinethickness{0.254mm}\path(310,-180)(360,-215) 
        \allinethickness{0.254mm}\path(310,-180)(350,-240) 
        \allinethickness{0.254mm}\path(175,-235)(230,-180) 
        \allinethickness{0.254mm}\path(180,-150)(230,-180) 
        \allinethickness{0.254mm}\path(195,-125)(230,-180) 
        \put(175,-235){{\ellipse*{10}{10}}} 
        \put(180,-150){{\ellipse*{10}{10}}} 
        \put(195,-125){{\ellipse*{10}{10}}} 
        \put(360,-215){{\ellipse*{10}{10}}} 
        \put(350,-240){{\ellipse*{10}{10}}} 
        \put(365,-125){{\ellipse*{10}{10}}} 
        \put(230,-171){\shortstack{${a}_2$}} 
        \put(290,-171){\shortstack{${a}_1$}} 
        \put(335,-126){\shortstack{${a}_3$}} 
        \put(320,-241){\shortstack{${a}_4$}} 
        \put(350,-206){\shortstack{${a}_5$}} 
        \put(160,-226){\shortstack{${a}_6$}} 
        \put(205,-131){\shortstack{${a}_7$}} 
        \put(170,-171){\shortstack{${a}_8$}} 
\end{picture}
}
\hspace{20pt}
\subfloat[]{
 \includegraphics[width=1in]{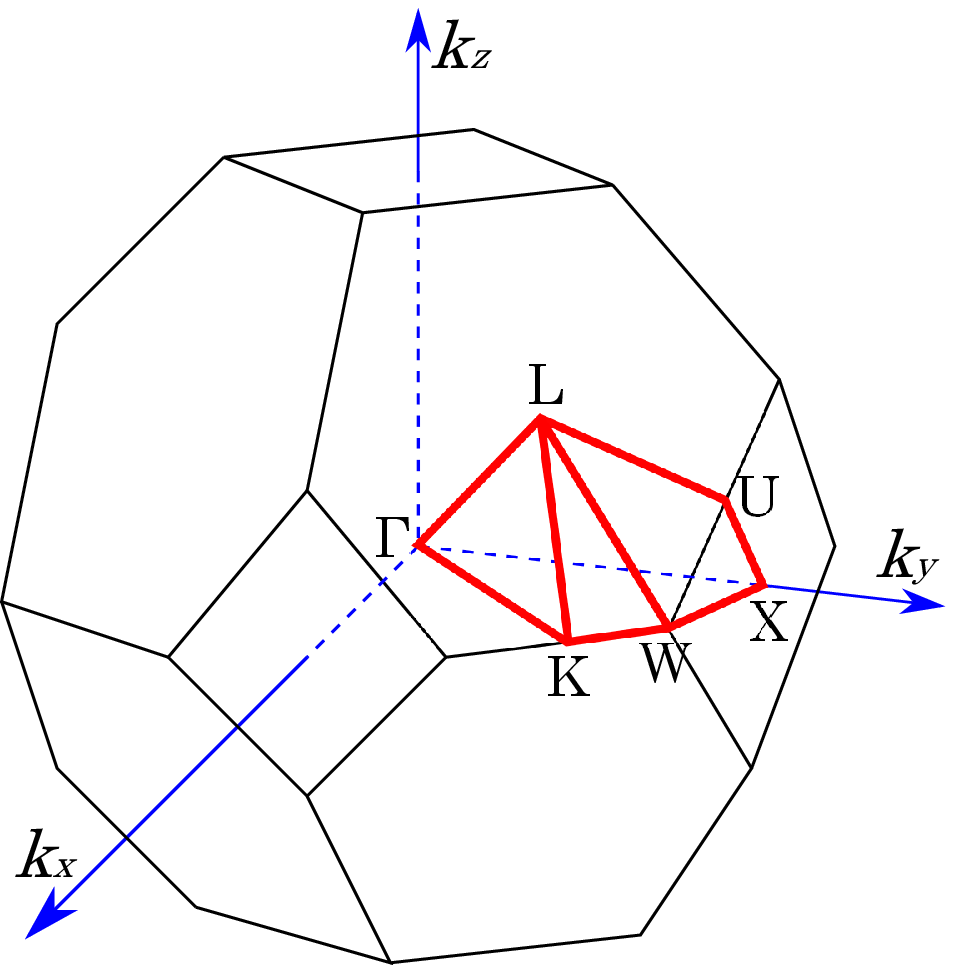} \label{ibz_dia}
}
         \caption{Tetrahedral lattice. (a) The unit cell. (b) The irreducible Brillouin zone \cite{Curtarolo2010}.}{\label{tetra_unit}}
\end{figure}

\subsubsection{Numerical result}
We consider an example for which all members are rods of radius $t$ and have the same uniform properties and are arranged in a regular tetrahedral lattice. The \rev2{numerical} computations are based on the  properties in Table \ref{tab4} and the path of the wave vector taken is along $\Gamma - {\text L}$ of the Brillouin zone shown in Fig.\ \ref{ibz_dia}.

\begin{table}[ht]
\caption{Tetrahedral lattice parameters.} 
\centering
\begin{tabular}{| c | c | c | c| c |}
  \hline
  $E$ (GPa) & $\nu$ & $\rho^V$ (kg/m$^3$)& $l$ (mm)& $t$ (mm) \\
    \hline  \hline
  70 & .33 & 2.7 $\cdot$ 10$^3$ & 10 &  .5 \\
  \hline
\end{tabular}
\label{tab4}
\end{table}
The dispersion curves in Fig.\ \ref{diamond_disp}(a)  were obtained by finding the smallest eigenvalue of a positive definite matrix, and plotting the corresponding wave number and frequency of the discretized grid where the smallest eigenvalue is smaller than $\epsilon$ (a small value). Figure \ref{diamond_disp} shows that the  dispersion curves  computed by the present simplified theory  agree well with those found using FEM. 
\rev2{As with the 2D hexagonal structure, the tetrahedral  lattice displays a broad frequency range with one-wave behaviour: 5 to 20 kHz.  The wave is almost non-dispersive, and isotropic in the long-wavelength regime on account of the symmetry of the lattice.}

 \begin{figure}[h!]  
\begin{center}
\subfloat[]{
 \includegraphics[width=2.5in]{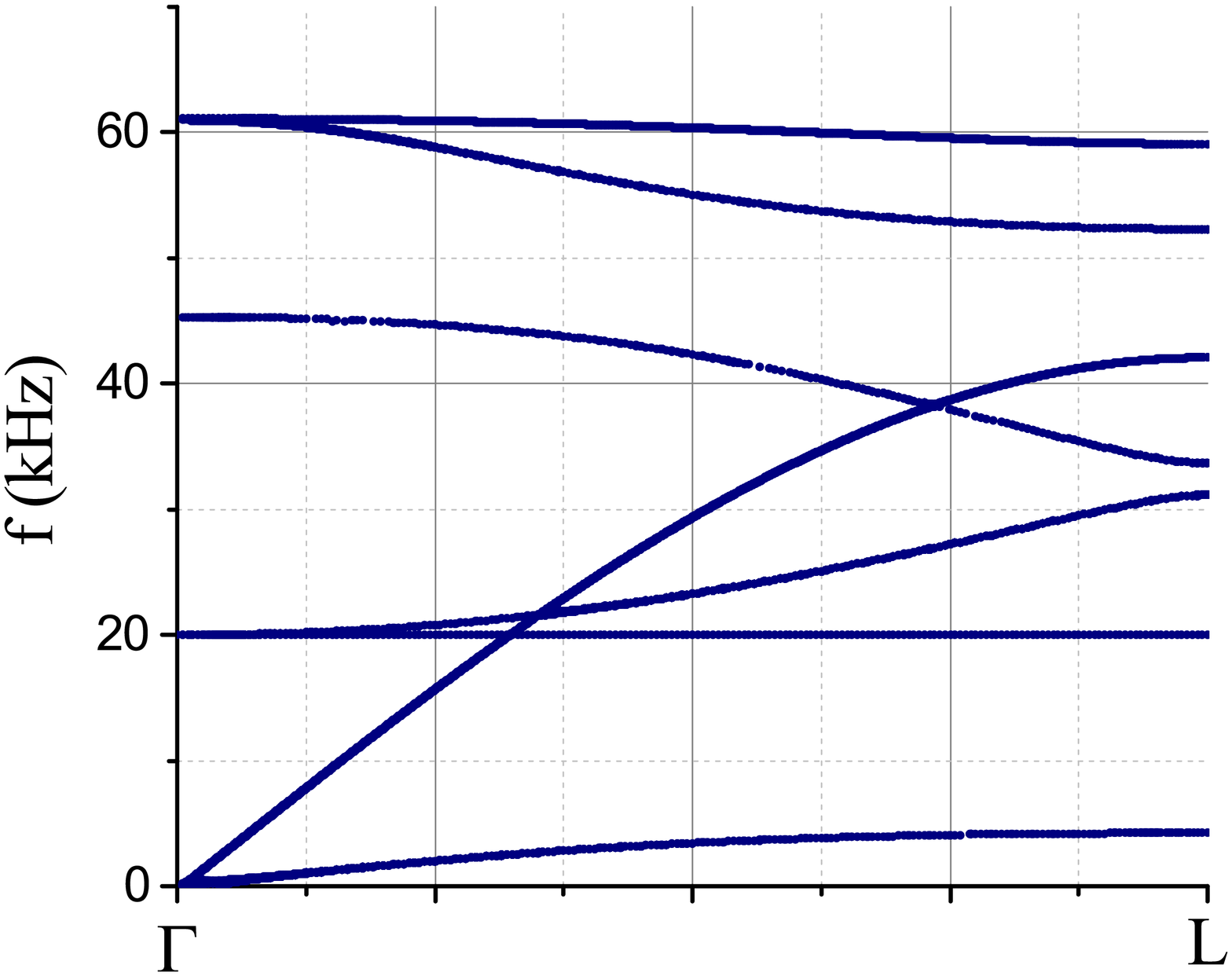}
}
\subfloat[]{
 \includegraphics[width=2.5in]{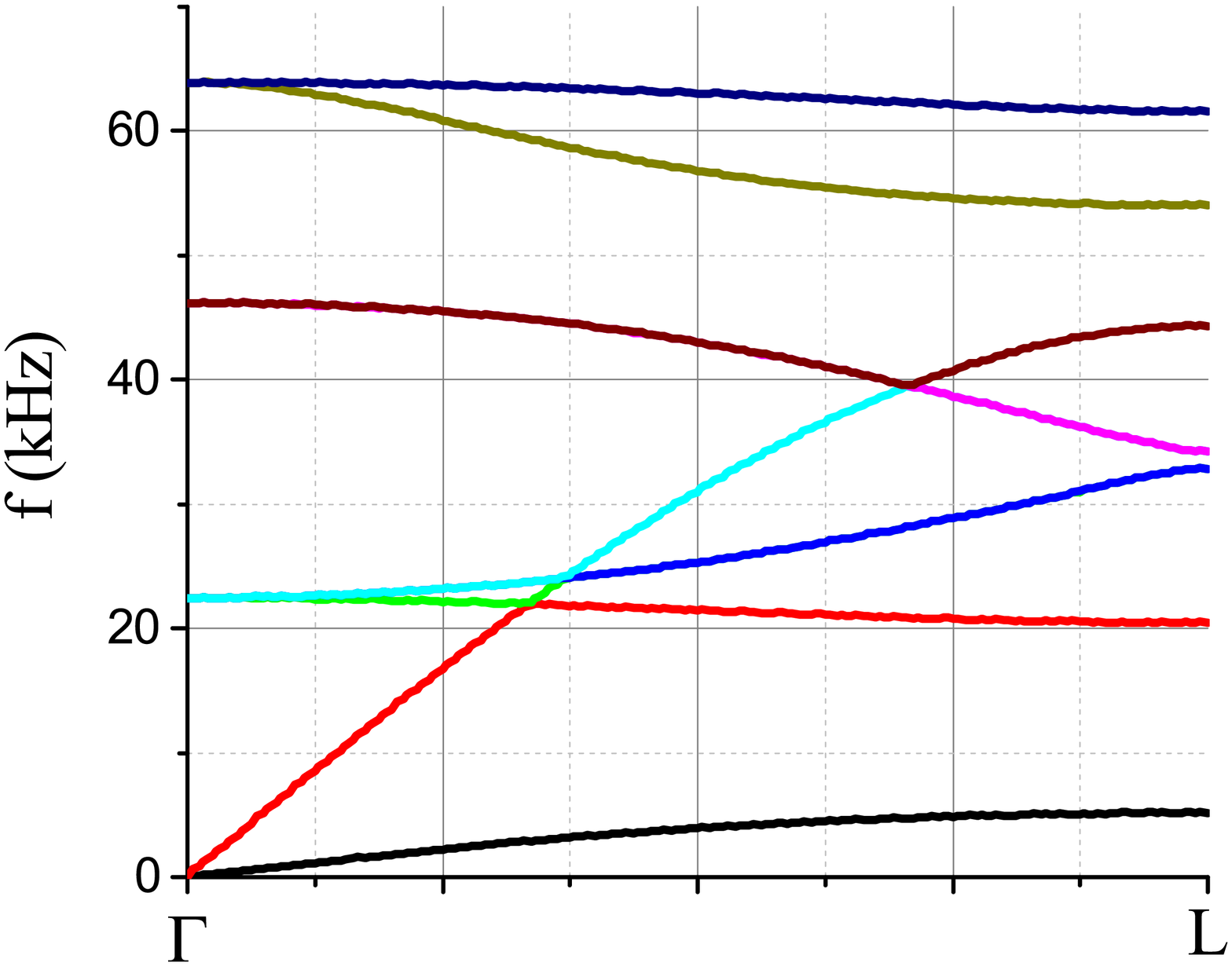}
}
\end{center}
\caption{Dispersion curves of the regular tetrahedral  lattice with properties in Table \ref{tab4}.  (a)  The first seven Floquet branches for wave-vector along the perimeter of the Brillouin zone.  (b)  Dispersion curves calculated using FEM (COMSOL).
\rev2{Note the clear  one-wave behaviour between about 5 and 20 kHz.}
}
 \label{diamond_disp}
\end{figure}

\section{Conclusions}\label{sec7}
Dynamic modeling of 2D and 3D lattices can be accurately  modeled using  a low order model with minimal degrees of freedom described by thin beam members. The dispersion relations for rectangular and cubic lattices have been  derived analytically by  imposing  the   Bloch-Floquet periodicity condition, yielding an Hermitian eigenvalue problem for the unknown frequencies. Numerical methods were used to compute the band-diagrams for hexagonal and tetrahedral lattices. The semi-analytical approach allowed us to extract the low frequency asymptotics. In particular, the closed-form explicit expressions for the Christoffel matrix in the quasistatic regime for rectangular, hexagonal and cubic lattices were presented. Numerical comparisons of wave dispersion diagrams with  FEM simulations  indicate that the beam model provides good accuracy for lower modes.  The semi-analytical nature of the present model makes it the natural extension of  purely static methods for periodic lattice structures, e.g. \cite{Norris14}. 
\rev2{It accurately predicts the one-wave behaviour in the  hexagonal and  tetrahedral lattices.  These particular structures are distinct  in that they provide effective in the long-wavelength limit, and hence quasi-acoustic wave effects in the one-wave regions.  By breaking the symmetry one can extend the scalar one-wave effect to display anisotropy, an important  subject for future investigation  with the semi-analytic  model.}
 In summary, our beam model provides a novel and fast approach to calculate the band-diagrams for 2D and 3D lattices. This semi-analytical method may prove useful in designing phononic crystals and pentamode structures. 

\section*{Acknowledgments}
AK was partially supported by the RSF project N\textsuperscript{\underline{o}}15-11-30007 and TRR 181 project.   X.S. acknowledges support  under ONR MURI Grant No. N000141310631. A.N.N. acknowledges support from Institut de M\'{e}canique et d'Ing\'{e}nierie, Universit\'{e} de Bordeaux.  The reviewers are thanked for providing suggestions that improved the paper.


\begin{thebibliography}{10}

\bibitem{Milton95}
G.~W. Milton and A.~V. Cherkaev.
\newblock Which elasticity tensors are realizable?
\newblock {\em J. Eng. Mat. Tech.}, 117(4):483--493, 1995.

\bibitem{Norris14}
A.~N. Norris.
\newblock Mechanics of elastic networks.
\newblock {\em Proc. R. Soc. A}, 470:20140522+, 2014.

\bibitem{Martinsson03d}
P.~G. Martinsson and A.~B. Movchan.
\newblock Vibrations of lattice structures and phononic band gaps.
\newblock {\em Q. J. Mech. Appl. Math.}, 56(1):45--64, 2003.

\bibitem{Phani06}
A.~S. Phani, J.~Woodhouse, and N.~A. Fleck.
\newblock Wave propagation in two-dimensional periodic lattices.
\newblock {\em J. Acoust. Soc. Am.}, 119(4):1995--2005, 2006.

\bibitem{Gonella08}
S.~Gonella and M.~Ruzzene.
\newblock Analysis of in-plane wave propagation in hexagonal and re-entrant
  lattices.
\newblock {\em J. Sound. Vib.}, 312(1-2):125--139, 2008.

\bibitem{WITTRICK1971}
W.~H. Wittrick and F.~W. Williams.
\newblock A general algorithm for computing natural frequencies of elastic
  structures.
\newblock {\em Q J Mechanics Appl Math}, 24(3):263--284, 1971.

\bibitem{Spadoni09}
A.~Spadoni, M.~Ruzzene, S.~Gonella, and F.~Scarpa.
\newblock Phononic properties of hexagonal chiral lattices.
\newblock {\em Wave Motion}, 46(7):435--450, 2009.

\bibitem{Leamy12}
M.~J. Leamy.
\newblock Exact wave-based {B}loch analysis procedure for investigating wave
  propagation in two-dimensional periodic lattices.
\newblock {\em J. Sound. Vib.}, 331(7):1580--1596, 2012.

\bibitem{Martinsson03a}
P.~G. Martinsson and I.~Babu\v{s}ka.
\newblock {Homogenization of materials with periodic truss or frame
  micro-structures}.
\newblock {\em Math. Models Methods Appl. Sci.}, 17(5):805--832, 2007.

\bibitem{Gonella08b}
S.~Gonella and M.~Ruzzene.
\newblock {Homogenization and equivalent in-plane properties of two-dimensional
  periodic lattices}.
\newblock {\em Int. J. Solids Struct.}, 45(10):2897--2915, 2008.

\bibitem{Colquitt11}
D.~J. Colquitt, I.~S. Jones, N.~V. Movchan, and A.~B. Movchan.
\newblock Dispersion and localization of elastic waves in materials with
  microstructure.
\newblock {\em Proc. R. Soc. A}, 467(2134):2874--2895, 2011.

\bibitem{Colquitt13}
D.~J. Colquitt, M.~J. Nieves, I.~S. Jones, N.~V. Movchan, and A.~B. Movchan.
\newblock {Localisation for a line defect in an infinite square lattice}.
\newblock {\em Proc. R. Soc. A}, 469(2150):20120579, 2013.

\bibitem{Williams1995}
F.~W. Williams and J.~R. Banerjee.
\newblock Free vibration of composite beams - {A}n exact method using symbolic
  computation.
\newblock {\em Journal of Aircraft}, 32(3):636--642, May 1995.

\bibitem{Maldovanbook09}
Martin Maldovan and Edwin~L. Thomas.
\newblock {\em Periodic Materials and Interference Lithography: For Photonics,
  Phononics and Mechanics}.
\newblock Wiley-VCH, 2009.

\bibitem{Norris11mw}
A.N. Norris and A.J. Nagy.
\newblock Metal {W}ater: A metamaterial for acoustic cloaking.
\newblock In {\em Proceedings of Phononics 2011, Santa Fe, NM, USA, May 29-June
  2}, pages 112--113, Paper Phononics--2011--0037, 2011.

\bibitem{Curtarolo2010}
W.~Setyawana and S.~Curtarolo.
\newblock High-throughput electronic band structure calculations: Challenges
  and tools.
\newblock {\em Comp. Mat. Sc.}, 49:299--312, 2010.

\end{thebibliography}

\end{document}